\documentclass[a4paper, 11pt]{article}
\usepackage{comment} % enables the use of multi-line comments (\ifx \fi) 
\usepackage[nointegrals]{wasysym}
\usepackage{fullpage} % changes the margin
\usepackage{graphicx}
\usepackage{cite}
\usepackage{multicol}
\usepackage[perpage]{footmisc} 
\usepackage{sectsty}
\sectionfont{\large}
\usepackage{graphicx}
\usepackage{subcaption}
\usepackage{bm}
\usepackage{float}
\usepackage{amsmath}
\usepackage[utf8]{inputenc}
\usepackage{amssymb}
\usepackage{mathrsfs}
\usepackage{wrapfig}
\usepackage{enumitem}
\usepackage[english]{babel}
\usepackage{comment}
\usepackage{bbold}  % identity with \mathbb{1}
\usepackage[usenames,dvipsnames]{xcolor} 
\usepackage{xcolor,hyperref}
\usepackage{setspace}
\usepackage{abstract}
\usepackage[font=footnotesize,labelfont=bf]{caption}
\usepackage{tablefootnote}

\usepackage[T1]{fontenc}
\usepackage[margin=0.94in]{geometry}
\def\Mpl{M_{\rm P}}

\usepackage{enumitem}
\usepackage{graphicx}
\usepackage{subcaption}
\usepackage{xcolor}
\usepackage{amssymb}

\numberwithin{equation}{section}

\begin{document}
\null\hfill IPMU25-0037 \\ 
\null\hfill YITP-25-86 \\
\vspace*{\fill}

\vspace*{\fill}
\begin{center}
    \LARGE\textbf{{ \textcolor{Black}{Accelerating Universe from Constraints} }}

    \normalsize\textsc{Anamaria Hell$^{1}$, Misao Sasaki$^{1,2,3,4}$}
\end{center}

\begin{center}
    $^{1}$ \textit{Kavli IPMU (WPI), UTIAS,\\ The University of Tokyo,\\ Kashiwa, Chiba 277-8583, Japan}
    \\
    $^{2}$ \textit{Center for Gravitational Physics and Quantum Information,\\ Yukawa Institute for Theoretical Physics, Kyoto University, Kyoto 606-8502, Japan}\\
     $^{3}$ \textit{Leung Center for Cosmology and Particle Astrophysics,\\ National Taiwan University, Taipei 10617, Taiwan}\\
  $^{4}$ \textit{ Asia Pacific Center for Theoretical Physics,\\ Pohang 37673, Korea}
\end{center}
\thispagestyle{empty} 

\renewcommand{\abstractname}{\textsc{\textcolor{Black}{Abstract}}}

\begin{abstract}
 We introduce a framework of constrained scalar fields that can give rise to cosmological evolution of the Universe independent of the values of the cosmological constant. Focusing on the simplest realization involving a scalar field with non-minimal coupling, we first study the analytical solutions in the Jordan frame. We show that such solutions  
 include evolution from a radiation dominated-like universe to an exponential expansion, as well as evolution that starts with super-Hubble expansion and then relaxes towards an exponential expansion of the Universe, while describing well-behaved scalar perturbations. We then also relate the theory to the Einstein frame, and analyze the corresponding accelerating solutions. We then consider the model in the presence of external matter. 
 We find that the matter does not affect the evolution of the universe if minimally coupled to the scalar field. 
 In other words, in the Jordan frame, in order to influence the evolution of the space-time, the matter should be non-minimally coupled to the constrained scalar, while it may be minimally coupled to gravity.
 We show that in this case, the solutions are similar to the free case, and, in addition, allow for the phantom-like equation of state. 
 Finally, we introduce the minimal frame, a frame in which the matter is minimally coupled to both gravity and the constrained scalar, and show that among other possibilities, the phantom-like equation of state can still be realized. 
  \end{abstract}
 
\vfill
\small

\noindent\href{mailto:anamaria.hell@ipmu.jp}{\text{anamaria.hell@ipmu.jp}}\\
\href{mailto:misao.sasaki@apctp.org}{\text{misao.sasaki@apctp.org}}
\vspace*{\fill}

\clearpage
\pagenumbering{arabic} 
\newpage

\tableofcontents
\newpage
\section{Introduction}

One of the most exciting and important puzzles in modern cosmology is -- \textit{What drives the acceleration of our Universe?} Such solutions have been known for over a century, starting initially with the de Sitter solution, found only two years after the introduction of Einstein's general relativity \cite{dS1917, dS1918}. Yet, despite numerous models that we face today, the substance that gives rise to the early- or late-time acceleration is still uncertain.

Inflationary models -- theories that account for the early acceleration of space-time are some of the most favored candidates to describe the Universe just moments after its emergence \cite{Starobinsky:1980te, Sato:1980yn, Guth:1980zm, Linde:1981mu, Brout:1979bd, Albrecht:1982wi, Linde:1983gd} (see also \cite{Martin:2013tda} for a comprehensive list of models). If the Universe went through such an accelerated stage, these models could explain the origin of the large-scale structure we see today through quantum fluctuations of the spacetime \cite{Chibisov:1982nx, Mukhanov:1981xt}. 
Moreover, the simplest models predict a flat universe, equipped with adiabatic,  nearly Gaussian primordial curvature perturbations having a nearly scale-invariant power spectrum, and the presence of long-wavelength primordial gravitational waves \cite{Starobinsky:1979ty, Kodama:1984ziu,Mukhanov:1985rz, Sasaki:1986hm }. 
Most of these theories are based on a single scalar field, which may be parametrized with an equation of state \cite{Mukhanov:2013tua, Achucarro:2022qrl}. 
They give rise to acceleration once the potential of the scalar field dominates over its kinetic energy, thus providing dynamics that are close to the dS Universe.
In addition to single scalar field models, those with multiple fields, non-minimal coupling, or those arising from modified theories of gravity have also become attractive, giving rise to interesting phenomena such as primordial black  hole formation (see eg. \cite{ Gong:2016qmq, Byrnes:2025tji, Wang:2025lti, Sasaki:2025vql} and references therein ).

In addition to the very early expansion of our Universe, one of the most important puzzles to date is to uncover what drives its late-time acceleration \cite{SupernovaSearchTeam:1998fmf, SupernovaCosmologyProject:1998vns}. The simplest explanation is that this acceleration is driven by the cosmological constant, an essential part of the standard cosmological model $\Lambda${CDM}. Another candidate is dark energy (DE) --  a substance that dynamically evolves in time, in contrast to the constant value of $\Lambda$. Indeed, recent observations seem to favor this scenario \cite{DESI:2024mwx, DESI:2025zgx}. 
It naturally arises in scalar field models such as quintessence theories \cite{Peebles:1987ek, Fujii:1982ms, Ford:1987de, Wetterich:1987fm, Ratra:1987rm, Chiba:1997ej, Ferreira:1997au, Caldwell:1997ii, Coble:1996te, Turner:1997npq}, and in k-essence models \cite{
Armendariz-Picon:2000nqq, Armendariz-Picon:2000ulo}, which contain a non-trivial kinetic term for the scalar, and non-minimally coupled scalar fields \cite{Brans:1961sx,  Wetterich:1987fk, Wetterich:1987fm, Sakai:1998rg, Chiba:1997ij, Amendola:1999qq, Uzan:1999ch, Chiba:1999wt, Bartolo:1999sq, Perrotta:1999am, Elizalde:2004mq, Nojiri:2005pu, Martin:2005bp, Gannouji:2006jm, Carroll:2006jn, Bean:2006up, Perivolaropoulos:2005yv, Amendola:2007nt,Hu:2007nk, Starobinsky:2007hu,Tsujikawa:2007xu,   Elizalde:2008yf, Tsujikawa:2008uc,  Bamba:2008hq, Motohashi:2010tb,  Gannouji:2010fc, Wetterich:2013jsa, Wetterich:2014gaa, Wolf:2025jed}. 
To date, numerous models for DE exist, taking place also in modified theories of gravity, which range from scalar-tensor theories, such as $f(R)$ gravities\footnote{It is natural to consider this theory as part of modified theories if matter is minimally coupled to gravity in the Jordan frame. However, one should note that it is also possible to write this theory in an Einstein frame, where it becomes equivalent to Einstein's gravity with a scalar field up to few exceptions \cite{Whitt:1984pd, Maeda:1988ab, Wands:1993uu, Hell:2023mph}. }, which formally correspond to the $\omega=0$ Brans-Dicke theory, to also much more complicated models (see eg.  \cite{ Brans:1961sx, Dvali:2000hr, Chiba:2003ir, Khoury:2003rn,  Copeland:2006wr, Sotiriou:2008rp,  Silvestri:2009hh, Nojiri:2010wj, Tsujikawa:2010sc,Tsujikawa:2010zza, DeFelice:2010aj,  Clifton:2011jh, Deffayet:2011gz, Kobayashi:2011nu, Charmousis:2011bf, Bamba:2012cp, Joyce:2014kja, Bull:2015stt, Takahashi:2015ifa, Nojiri:2017ncd,Aoki:2018brq, CANTATA:2021asi,  Odintsov:2023weg, DiValentino:2025sru  }).

Yet,
its nature brings several puzzles, making it thus one of the greatest mysteries in fundamental physics \cite{Weinberg:1988cp, Sahni:1999gb, Peebles:2002gy, Padmanabhan:2002ji, Sahni:2004ai, Copeland:2006wr, Sahni:2006pa}. 
The first is the magnitude of DE -- observations tell us that it is smaller than the theoretical prediction by 120 orders of magnitude, thus giving rise to the cosmological constant problem.
The second puzzle that arises with DE is its behavior at low redshifts. The data suggests that the associated equation of state $w$ is smaller than $-1$, thus violating the null-energy condition, which is called the phantom equation of state\footnote{See also \cite{Shlivko:2025fgv} for the discussion on the parametrization for certain type of DE models.}.
For most of the models, it indicates the presence of ghost-like modes which are pathological, as they make the energy unbounded from below. Despite this, some models, such as ghost inflation, Hordenski theory, generalized Proca, or stealth solutions in k-essence and string-inspired quintessence can accommodate such equation of state \cite{Horndeski:1974wa,Arkani-Hamed:2003pdi, Mukohyama:2005rw, Creminelli:2006xe, DeFelice:2016uil, Alberte:2016izw, Anchordoqui:2025fgz}. Notably, some models of $f(R)$ gravity might also account for the phantom-like equation of state \cite{Motohashi:2010zz, Motohashi:2010tb, Motohashi:2011wy}.

Given the increasing precision in cosmological observations, it is important to explore theories that might accommodate both the interesting evolution of the dark energy, accounting for its small value, as well as the early-time acceleration of our Universe, while being theoretically consistent. In this work, we will explore one of such roads by considering a new framework, where the dynamics of the Universe will arise due to the constrained scalar fields -- fields lacking a standard kinetic term -- that are coupled to curvature in such a way that the cosmological background evolution is independent of the values of cosmological constant.  In particular, here, we will focus on the simplest possible realization of such an idea with a constrained scalar that is non-minimally coupled to the Ricci scalar.

By first examining its analytical structure in the absence of matter, we find that this model admits solutions with intriguing properties, that could be important for both early- and late-time acceleration of the universe, even in the absence of matter.
Once we include a matter field, we have to decide how it couples to gravity as well as to the constrained scalar. 
We find that this gives rise to a new concept -- the minimal frame -- which is the frame in which the matter is minimally coupled to both gravity and the constrained scalar and thus gives a natural choice of the physical frame.
Here, we emphasize that the purpose of this work is not to construct an observationally viable solution, but rather to introduce a new theoretical framework and to illustrate its characteristic features through the analysis of the simplest realizations. Through suitable extensions, we expect that the current framework may accommodate the actual evolution of our Universe.

The paper is organized in the following way. First, in Section 2, we will introduce the framework of the constrained scalar by presenting its simplest realization, which will be the main focus of this work. We will study this theory in the absence of external in the Jordan frame, and extend the analysis also to the Einstein frame in Section 3, and analyze the perturbations in Section 4. 
In Section 5 we will take into account the possible coupling of the external matter, and discuss the subtleties and surprises that arise in this case. 
In Section 6, we will introduce the minimal frame, which defines the physical frame. Finally, Section 7 is devoted to conclusion and discussion.

\textsc{\textbf{Conventions:} } Throughout the paper, we will use the metric signature $(- + + +)$. 
In addition, we will be working in several different frames -- the Jordan frame, the Einstein frame, and the minimal frame. 
In each of them, the dot, such as $\dot{Q}$ denotes a derivative with respect to the time of that frame.
The derivative with respect to the spatial coordinate $x^i$ will be denoted by a comma, such as $Q_{,i}$.

\section{The free theory of a constrained scalar}
When we consider field theories that could be relevant for particle physics or cosmology, we start by writing down their kinetic terms. 
However, it is also sensible to ask if the kinetic terms are always necessary. As we will see in this work, it is possible to realize interesting cosmological evolution even if a scalar field has no standard kinetic term, if it is non-minimally coupled to the spacetime curvature. 
In particular, to illustrate this idea, let us consider a scalar-tensor theory with the action in the Jordan frame given by  
\begin{equation}\label{JframeAction}
    \begin{split}
        S=\int d^4x\sqrt{-g}&\left[\frac{\Mpl^2}{2}\left(R-2\Lambda\right)-\frac{m^2}{2}\sigma^2+\frac{\xi}{2}R\sigma^2\right]+S_m\,.
    \end{split}
\end{equation}
Here, $R$ is the Ricci scalar, $\Lambda$ is a cosmological constant, $\xi$ and $m$ are constants, and $\sigma$ is the scalar field. In addition, $S_m$ describes the matter action. 
We notice that the scalar field $\sigma^2$ in this theory is very special -- it lacks a standard kinetic term.
As a result, if $S_m$ does not involve $\sigma$, the variation of the action with respect to $\sigma$ leads to the constraint,
\begin{equation}
    \sigma\left(\xi R-m^2\right)+\frac{\delta S_m}{\delta\sigma}=0\,.
\end{equation}

First, let us assume that $S_m$ is independent of $\sigma$, or simply ignore the matter field. In this case, the above equation admits two solutions. 
The first possibility is that the scalar field vanishes, resulting in the standard Einstein gravity with a cosmological constant. Another more interesting case is when the scalar is non-vanishing. In this case the Ricci scalar is constrained to be a constant, $R=m^2/\xi$.
Clearly, $R$ is independent of the value of $\sigma$ as well as of the value of the cosmological constant $\Lambda$.
In passing, we note that if the mass term is replaced by a potential $V(\sigma)$ as in the general massive dilaton, the Ricci scalar will be determined by the derivative of the potential $V'(\sigma)$, but it will still be independent of $\Lambda$.

Having the curvature being independent of the value of the cosmological constant is intriguing: It opens a novel framework for the scalar field theories, which can potentially account for the early- and late-time acceleration of our Universe, with possible generalizations of the previous action such as 
\begin{equation}\label{JframeActionGeneral}
    \begin{split}
        S=\int d^4x\sqrt{-g}&\left[\frac{\Mpl^2}{2}\left(R-2\Lambda\right)-\frac{m^2}{2}g(\sigma)+\frac{\xi}{2}f(R)g(\sigma)\right]+S_m\,, 
    \end{split}
\end{equation}
where $g(\sigma)$ is an arbitrary function of $\sigma$, and
$f(R)$ is an arbitrary function of the curvature that may involve not only the Ricci scalar but also general combinations of the Riemann tensor. 
%However, it is therefore natural to be suspicious if such theories might be cosmologically relevant. 
However, in this work, we will focus on the simplest case described by the action (\ref{JframeAction}). 
While this simplest case might not necessarily realize an observationally viable cosmological scenario, it will nevertheless be interesting from the point of theoretical consistencies -- whether it gives theoretically healthy background solutions.

One should note that the usual models of a scalar field with non-minimal coupling to gravity involve a standard kinetic term for the scalar field, with the most well-known case being the general Brans-Dicke theory \cite{Brans:1961sx} and its extensions via the scalar-field potential \cite{Sakai:1998rg, Chiba:1997ij, Amendola:1999qq, Uzan:1999ch, Chiba:1999wt, Bartolo:1999sq, Perrotta:1999am}. 
However, there is also a special case of Brans-Dicke theory with $\omega=0$, which is also known as the massive dilaton gravity (MDG) \cite{OHanlon:1972xqa, Wands:1993uu, Sotiriou:2008rp},
\begin{equation}
    S_{\omega=0}=\int d^4x\sqrt{-g}\left[\chi R-V(\chi)\right]\,.
\end{equation}
Notably, with the identifications, $\chi=f'(\Phi)$ and $V(\chi)=\Phi f'(\Phi)-f(\Phi)$,
this theory is equivalent to the $f(R)$ gravity \cite{Teyssandier:1983zz}.

The minimal realization of our framework of a constrained scalar field (\ref{JframeAction}) can also be related to the above theory.  Specifically, if we start from the MDG with linear potential,
\begin{equation}
     S_{BD}=\int d^4x\sqrt{-g}\left[\chi R-\lambda\chi\right],
\end{equation}
 and substitute
\begin{equation}
    \chi=\frac{\Mpl^2}{2}+\frac{\xi\sigma^2}{2}\,,
\end{equation}
with the identifications,
\begin{equation}
    \xi\Lambda=2m^2, \qquad \lambda=\frac{\Lambda}{2}\,,
\end{equation}
we recover the initial action (\ref{JframeAction}). {Here, one should still keep in mind that the above relation holds only if mass, coupling constant and the cosmological constant are connected in a particular way, which is apriori not assumed in (\ref{JframeAction}).  }

In this section, we will explore the cosmological solutions of the theory (\ref{JframeAction}) in the Jordan frame. For simplicity we set $S_m=0$.
The variation of the action  with respect to the metric gives
\begin{equation}
    R_{\mu\nu}\left(\frac{\Mpl^2}{2}+\frac{\xi}{2}\sigma^2\right)-\frac{1}{2}g_{\mu\nu}\left[R\left(\frac{\Mpl^2}{2}+\frac{\xi}{2}\sigma^2\right)-M_{\rm P}^2\,\Lambda-\frac{m^2}{2}\sigma^2\right]=\frac{\xi}{2}\left[\nabla_{\mu}\nabla_{\nu}\sigma^2-g_{\mu\nu}\Box\sigma^2\right]\,,
\label{Gmn}
\end{equation}
and that with respect to $\sigma$ gives
\begin{equation}
    \sigma(\xi R- m^2)=0\,.
    \label{sigmaconstraint}
\end{equation}

Let us assume a spatially flat, homogeneous and isotropic background,
\begin{equation}
    ds^2=-dt^2+a(t)^2\delta_{ij}dx^idx^j,
\end{equation}
where $a(t)$ is the scale factor. 
In addition, we will assume that the scalar field takes a non-vanishing background value $\sigma (t)$. 
The time-time component of Eq.~(\ref{Gmn}) gives the Hamiltonian constraint, or the Friedmann equation in the case of Einstein gravity,
\begin{equation}
 \left(-6 \xi  \,H^{2}+m^{2}\right) \sigma^{2}-12 \dot{\sigma} \sigma  H \xi +2 \Mpl^{2} \left(-3 H^{2}+\Lambda \right)=0,
\end{equation}
where the dot denotes the derivative with respect to the (Jordan frame) time. 
The trance of the space-space components of Eq.~(\ref{Gmn}) gives
\begin{equation}\label{acceq0}
   \left(-6 \xi  \,H^{2}-4 \xi  \dot{H} +m^{2}\right) \sigma^{2}-4 \xi  \left(2 \dot{\sigma}  H +\ddot{\sigma}\right) \sigma -6 \Mpl^{2} H^{2}-4 \dot{\sigma}^{2} \xi +2 \Mpl^{2} \left(\Lambda -2 \dot{H} \right)
=0\,.
\end{equation}
Finally, Eq.~(\ref{sigmaconstraint}) gives the constraint,
\begin{equation}
    \sigma  \left(-12 \xi  \,H^{2}-6 \xi  \dot{H} +m^{2}\right)=0\,.
\end{equation}
Since $\sigma\neq0$, we %can divide the above equation by $\sigma$, and 
obtain an equation that involves only the Hubble parameter,
\begin{equation}
    -12 \xi  \,H^{2}-6 \xi  \dot{H} +m^{2}=0\,.
    \label{scalarconstr}
\end{equation}
Among the above three equations, only two are necessary to solve for the constrained scalar and the Hubble parameter. In particular, we can notice that Eq. (\ref{acceq0}) is automatically satisfied by combining the remaining two equations, just the case of Einstein gravity.

To find cosmological solution, we solve the following two equations: 
\begin{equation}\label{systemHsigma}
    \begin{split}
      &\dot{H}(t)=\frac{12 \xi  \,H(t)^{2}-m^{2}}{-6 \xi},%\qquad\text{and},
      \\
    &\dot{\Theta}(t)=\frac{1}{6H\xi}\left[(m^2-6\xi H^2)\Theta(t)+2\Mpl^2(\Lambda-3H^2)\right]; \qquad \Theta(t)\equiv\sigma^2. 
    \end{split}
\end{equation}
Notably, the Hubble parameter does not depend neither on the scalar field nor on the cosmological constant. Therefore, in the following, we will first study its solutions. Then, we will use the values of the Hubble parameter to further analyze the behavior of the scalar field. 

\subsection{The Hubble parameter}\label{section::HubP}
By substituting 
\begin{equation}
    H=\frac{\dot{y}(t)}{2y(t)}\,,
\end{equation}
the first equation in (\ref{systemHsigma}) becomes
\begin{equation}
    \Ddot{y}(t)=\frac{m^2}{3\xi}y(t)\,. 
\end{equation}
Its general solution is given by:
\begin{equation}
    y(t)=\alpha\cosh\frac{mt}{\sqrt{3\xi}}+\beta\sinh\frac{mt}{\sqrt{3\xi}},
\end{equation}
where $\alpha$ and $\beta$ are integration constants. Therefore, the general solution for the Hubble parameter is given by
\begin{equation}
    H(t)=\frac{m}{2\sqrt{3\xi}}\frac{\alpha\sinh\gamma+\beta\cosh\gamma}{\alpha\cosh\gamma+\beta\sinh\gamma},\qquad\text{where}\qquad\gamma=\frac{mt}{\sqrt{3\xi}}\,.
\end{equation}
\textcolor{black}{In this work, we will only focus on the expanding solutions. } Depending on the values of the integration constants, we will have four distinct interesting cases, which we will analyze below. 

\subsubsection*{ \color{Black}{$\diamond$} \textbf{{\textcolor{Black}{Case 1:}}} {Super-Hubble to exponential expansion} }
Let us first set $\beta=0$. In this case, the solution simplifies to:
\begin{equation}\label{satoa}
    H(t)=\frac{m}{2\sqrt{3\xi}}\tanh\frac{mt}{\sqrt{3\xi}}
\end{equation}
 The above solution describes an accelerated expansion (with $t_0=0$), with a curious feature -- the Hubble parameter grows with time, while in a usual case, it decreases, and thus gives rise to the \textit{super-Hubble expansion}. The scale factor is found to be
 \begin{equation}
     a(t)=a_0{\cosh^{1/2}\frac{mt}{\sqrt{3\xi}}}\,; \quad t>0\,.
 \end{equation}
 One can notice that an equivalent solution can be obtained by replacing $m^2=-\mu^2$ and $\xi=-\zeta$ at the same time if both $m^2$ and $\xi$ are negative. 
 
 At $t\sim 0$, the Hubble parameter can be approximated as
\begin{equation}
    H(t)\sim\frac{m^2 }{6\xi}t \,.
\end{equation}
Hence the scale factor starts from a finite value, and grows super-exponentially with time at early times, 
\begin{equation}
    a(t)\sim a_0e^{\frac{m^2t^2}{\xi}}\,.
\end{equation}
At late times, $m\,t/\sqrt{3\xi}\gg 1$, the Hubble parameter tends to a constant, 
\begin{equation}
   H\to   H_{lt}=\frac{m}{2\sqrt{3\xi}}\,, 
\end{equation}
resulting in the standard exponential time dependence of the scale factor,
\begin{equation}
    a(t)\sim e^{H_{lt}t}\,.
\end{equation}

We should note that the above theory also admits another solution if $m^2>0$ but $\xi=-\zeta<0$. Then, the solution for the Hubble parameter is given by
\begin{equation}
    H = 
\frac{m}{2 \sqrt{3\zeta}}\tan \! \left(\frac{m  t}{\sqrt{3\zeta}}\right)\,,
\end{equation}
with the corresponding scale factor given by
\begin{equation}
    a = 
 \frac{a_0}{\cos^{1/2}\left(\frac{m \,t}{\sqrt{3\zeta}}\right)}\,;\quad 0<\frac{m\,t}{\sqrt{3\zeta}}<\frac{\pi}{2}\,.
\end{equation}
Thus, in this case, the universe expands super-exponentially all the time, and ends up with a big-rip singularity where the Hubble parameter diverges.

\subsubsection*{\color{Black}{$\diamond$} \textbf{{\textcolor{Black}{Case 2: }}}Radiation dominated to exponential expansion }
Let us now set $\alpha=0$ instead. In this case, we find
\begin{equation}\label{Hcase2}
    H(t)=\frac{\sqrt{3}m}{6\sqrt{\xi}}\coth\frac{mt}{\sqrt{3\xi}}\,.
\end{equation}
The scale factor is integrated to be
\begin{equation}
     a =a_0\,{\sinh^{1/2}\left(\frac{m \,t}{\sqrt{3\xi}}\right)}\,;\quad t>0\,.
\end{equation}

At $t=0$, this solution behaves like a radiation-dominated universe,
\begin{equation}
 H(t)\sim\frac{1}{2t}\,,~a(t)\propto t^{1/2}\,,
\end{equation}
while at late times, the Hubble parameter tends to a constant, 
\begin{equation}
 H\to     H_{lt}=\frac{m}{2\sqrt{3\xi}}\,,
\end{equation}
and thus describes an accelerating Universe. 

In contrast to the previous case which starts with a super Hubble expansion and evolves to a stage with an almost constant Hubble parameter, this case describes a universe which starts with a phase that looks exactly like radiation dominance, and eventually settles down to an exponential expansion phase with a constant Hubble parameter.
This result is intriguing, as we have not added any matter, but rather studied just the free theory. 
It indicates that, with slight modifications or generalizations, it may lead to a compelling dark energy mechanism, where the space-time itself mimics a universe filled with radiation and ultimately ends up with an exponential expansion.

\subsubsection*{\color{Black}{$\diamond$} \textbf{{\textcolor{Black}{Case 3:}}} Exponential expansion}
In addition to the previous two cases in which the Hubble parameter is evolving, it is possible to also have constant solutions for all times. This corresponds to setting $\alpha=\beta$. Then, we can easily find that the Hubble parameter is constant,
\begin{equation}
      H(t)=\frac{m}{2\sqrt{3\xi}}\,, ~ a(t)=a_0\exp\left[\frac{m\,t}{2\sqrt{3\xi}}\right]\,.
\end{equation}

\subsubsection*{ \color{Black}{$\diamond$} \textbf{{\textcolor{Black}{Case 4:}}} The general parameters}
The previous three cases were special, as we had picked the parameters $\alpha$ and $\beta$. In addition to them, we can also have the general case, in which $\alpha\neq\beta\neq0$.  
Then, the Hubble parameter is given by: 
\begin{equation}
    H(t)=\frac{m}{2\sqrt{3\xi}}\frac{\alpha\sinh\gamma+\beta\cosh\gamma}{\alpha\cosh\gamma+\beta\sinh\gamma},\qquad\text{where}\qquad\gamma=\frac{mt}{\sqrt{3\xi}},
\end{equation}
and the scale factor satisfies: 
\begin{equation}
    a(t)=a_0\sqrt{\alpha\cosh\gamma+\beta\sinh\gamma}\,, 
\end{equation}
where the range of $t$ (=$\gamma$) is such that the argument of the square root is positive, which depends on the values of $\alpha$ and $\beta$. The first two cases correspond to the case of $\alpha\neq0$, $\beta=0$ and $\alpha=0$, $\beta\neq0$, respectively. 
Although we do not explicitly show, one can easily extend the above to the cases of $m^2<0$ and/or $\xi<0$.
Throughout the rest of the paper, we focus on the case $m^2>0$ and $\xi>0$, unless otherwise specified.
We present the behavior of the Hubble parameter for several examples in Fig.~\ref{HThetaJordan}. 

\subsection{The scalar field}
We have seen that the evolution of the Hubble parameter and the scale factor, which are entirely determined by the mass of the constrained scalar and its non-minimal coupling. 
The values of the Hubble parameter then determine the evolution of the constrained scalar, the equation of which we recapitulate,
\begin{equation}\label{eomTh}
\begin{split}
    &\dot{\Theta}(t)=\frac{1}{6H\xi}\left[(m^2-6\xi H^2)\Theta(t)+2\Mpl^2(\Lambda-3H^2)\right], \qquad \Theta(t)=\sigma^2.
    \end{split}
\end{equation}
In this subsection, we will study the evolution of the scalar field for various different behaviors of the Hubble parameter. 

Before proceeding to the analysis, we note that we may regard $\Theta$ as the
fundamental variable instead of $\sigma$, as there is no kinetic term of $\sigma$ in the Lagrangian. In this case, one may allow negative values of $\Theta$. 
The lower bound is given by the condition that the theory is healthy. 
The action \eqref{JframeAction} indicates that the theory is well-defined as long as $\xi\Theta+\Mpl^2>0$. 
In fact, we will show in Sec.~\ref{sec:dof}, the no-ghost condition is guaranteed if this inequality is satisfied. 
Although we assume the positivity of $\Theta$ in the following analysis, unless otherwise mentioned, it is worth noting this point.

\subsubsection*{\color{Black}{$\diamond$} \textbf{{\textcolor{Black}{Case 1: Super-accelerated to exponential expansion}}  } }\label{Case1SE}
First, consider the case $\beta=0$, for which the Hubble parameter is given by
\begin{equation}\label{JC1Hub}
    H(t)=\frac{m}{2\sqrt{3\xi}}\tanh\frac{mt}{\sqrt{3\xi}}\,.
\end{equation}
and describes the evolution from a super-accelerated to an exponentially expanding phase.
In this case, the equation of motion for the constrained scalar can be also solved exactly. To solve it, it is convenient to introduce a new variable,
\begin{equation}
    u=\sinh{\frac{mt}{\sqrt{3\xi}}}\,.
\end{equation}
Then, Eq.~(\ref{eomTh}) becomes
\begin{equation}
    \frac{d\Theta}{du}=\frac{1}{u}\left[\left(1-\frac{1}{2}\frac{u^2}{1+u^2}\right)\Theta+\frac{2\Mpl^2}{m^2}\left(\Lambda-\frac{m^2}{4\xi}\frac{u^2}{1+u^2}\right)\right].
\end{equation}

The solution to the above equation is given by
\begin{equation}\label{Case1Sol}
        \Theta(u)=\frac{1}{2}\Mpl^2 \left(\frac{2\Lambda \xi
   -m^2}{m^2 \xi }\,{}_2F_1\left(\frac{1}{2},\frac{3}{4};\frac{3}{2};-u^2\right)
   \frac{u^2}{(u^2+1)^{1/4}} -\frac{4\Lambda}{m^2}
   +c_\Theta \frac{u}{ (u^2+1)^{1/4}}\right)\,,
\end{equation}
where $_2F_1$ is the hypergeometric function, and $c_{\Theta}$ is a constant of integration. 
In the limit $t\to0$, we have 
\begin{equation}
    \Theta\to \frac{M_{\rm P}^2}{2}
    \left(-\frac{4\Lambda}{m^2}+c_\Theta\, u+O(u^2)\right)\,.
    \label{Thetazero}
\end{equation}
Hence if we require the solution to be valid for $t>0$, $\Lambda$ must be non-positive.
Otherwise, one has to restrict the validity range of $t$ such that the condition 
{$\Theta>0$} is guaranteed. 
In addition, as one can see from \eqref{Thetazero}, taking $c_\Theta>0$ is necessary to ensure the positivity of $\Theta$ at $u\lesssim 1$. 

At $u\gg1$, the asymptotic form of the above gives 
\begin{equation}\label{asympInfC1}
    \Theta\to \frac{\Mpl^2}{2}\left(\sqrt{\pi }\frac{\Gamma
   \left(\frac{1}{4}\right)}{\Gamma \left(\frac{3}{4}\right)}\frac{  \left(2 \Lambda \xi -m^2\right)}{2 m^2
   \xi  }+c_{\Theta}\right) \sqrt{u}+\mathcal{O}\left(1\right)\,.
\end{equation}
Thus, the constant of integration $c_\Theta$ must be chosen so that the inequality, 
\begin{equation}
    c_\Theta> \sqrt{\pi }\frac{\Gamma
   \left(\frac{1}{4}\right)}{\Gamma \left(\frac{3}{4}\right)}\frac{  \left(m^2-2 \Lambda \xi \right)}{2 m^2
   \xi  }\,,
\end{equation}
is satisfied, in order to maintain $\Theta>0$ at asymptotic future.

\begin{figure}[h!]
    \centering

    \begin{minipage}{0.45\textwidth}
        \centering
        \includegraphics[width=\linewidth]{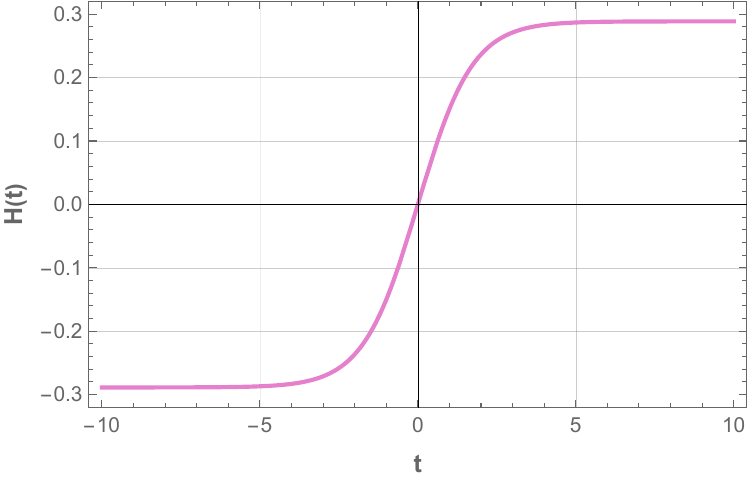}
    \end{minipage}
    \hfill
    \begin{minipage}{0.45\textwidth}
        \centering
        \includegraphics[width=\linewidth]{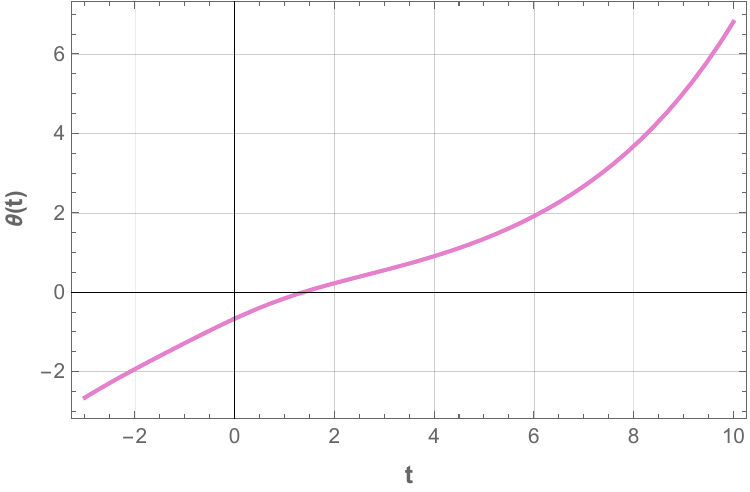}
    \end{minipage}

    \caption{{The evolution of the Hubble parameter and the constrained scalar for the parameters, $\alpha=1$, $\beta=0$, $m=1$, $\xi=1$, $\Lambda=1/3$ and $c_{\Theta}=1$ in the reduced Planck units $\Mpl=1$. 
    One should note that the constrained scalar $\sigma$ vanishes at $t>0$. 
    Note also that the $\sigma^2=\Theta<0$ region may be allowed as long as $\Theta>-\Mpl^2/\xi$.}
} 
    \label{Case1JThetaH}
\end{figure}

In Fig.~\ref{Case1JThetaH}, we demonstrate the behavior of the constrained scalar along with the value of the Hubble parameter. 

Overall, we see that the simple constrained scalar theory provides a feature that may potentially have interesting implications for the early Universe. Namely, the possibility that the Universe starts with an super-acceleration phase, that then transitions into the inflationary stage.
Here, it is important to note that this simplest theory lacks a graceful exit from inflation, which must eventually be incorporated if it is to be developed into a realistic theory.

\subsubsection*{\color{Black}{$\diamond$} \textbf{{\textcolor{Black}{Case 2: Radiation dominated to exponential expansion}}  } }\label{Case2RDexp}
Let us now set $\alpha=0$.  In this case, we have
\begin{equation}\label{case2H}
    H(t)=\frac{m}{2\sqrt{3\xi}}\coth\frac{mt}{\sqrt{3\xi}}\,.
\end{equation}
Similar to Case 1, setting $u=\sinh(mt/\sqrt{3\xi})$, the scalar equation \eqref{eomTh} now becomes
\begin{equation}
    \frac{d\Theta}{du}=\frac{u}{u^2+1}\left[\left(\frac{u^2-1}{2u^2}\right)\Theta+\frac{2\Mpl^2}{m^2}\left(\Lambda-\frac{m^2}{4\xi}\frac{u^2+1}{u^2}\right)\right].
\end{equation}
This can be solved exactly to give
\begin{equation}\label{solCase2}
   \begin{split}
     \Theta(u)&=\frac{\Mpl^2}{m^2 \xi } \sqrt{u^2+1}  \left[\,
   _2F_1\left(\frac{1}{4},\frac{1}{2};\frac{5}{4};-u^2\right)
   \left(2 \Lambda \xi-m^2 \right)-\frac{2 \Lambda \xi
   }{\sqrt{u^2+1}}+c_\Theta\frac{1}{\sqrt{u}} \right]\,.
   %+c_{\Theta} \sqrt{\frac{x^2+1}{x}},
   \end{split}
\end{equation}

\begin{figure}[h!]
    \centering

    \begin{minipage}{0.45\textwidth}
        \centering
        \includegraphics[width=\linewidth]{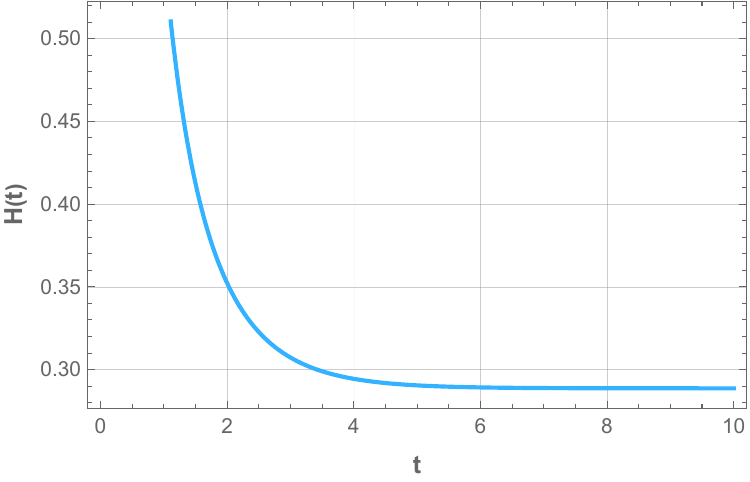}
    \end{minipage}
    \hfill
    \begin{minipage}{0.45\textwidth}
        \centering
        \includegraphics[width=\linewidth]{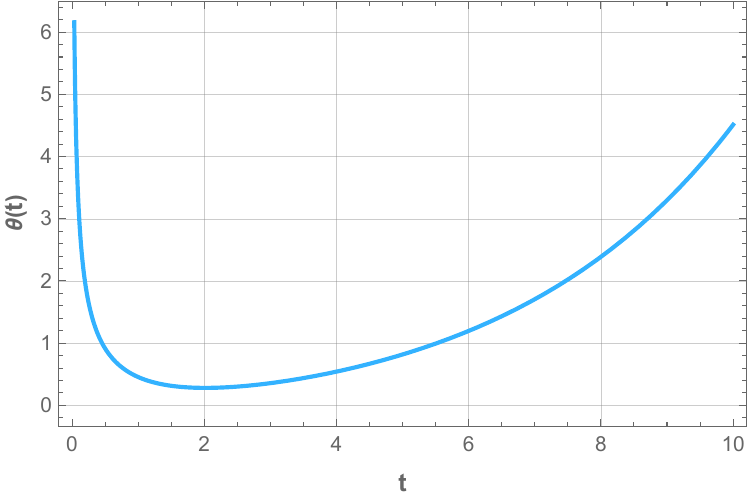}
    \end{minipage}

    \caption{{The evolution of the Hubble parameter and the constrained scalar for parameters: $\alpha=1$, $\beta=0$, 
   $m=1$, $\xi=1$, $\Lambda=1/3$ and $c_{\Theta}=1$ in the reduced Planck units $\Mpl=1$.  }
} 
    \label{Case2Jordan}
\end{figure}

At $t\to0$, or, equivalently $u\to 0$, we find that the above solution becomes
\begin{equation}
    \Theta\to\frac{c_{\Theta}}{\sqrt{u}}
    -\frac{M_{\rm P}}{\xi}+
    \mathcal{O}\left(\sqrt{u}\,\right)\,.
\end{equation}
Thus, we have to require $c_{\Theta}>0$ for the above solution to take positive values at early times. 
At late times, $u\gg 1$, we find find the following asymptotic form of (\ref{solCase2}):
\begin{equation}
    \Theta\to  \left(\frac{\Gamma \left(\frac{1}{4}\right) \Gamma
   \left(\frac{5}{4}\right)}{\sqrt{\pi } }\frac{\Mpl^2  \left(2 \Lambda \xi -m^2\right)}{m^2 \xi
   }+c_\Theta\right)\sqrt{u}+O(1)\,.
\end{equation}
Thus, for the constrained scalar to be positive at late times, we require
\begin{equation}
    c_\Theta>\frac{\Gamma \left(\frac{1}{4}\right) \Gamma
   \left(\frac{5}{4}\right)}{\sqrt{\pi } }\frac{\Mpl^2  \left(m^2 -2 \Lambda \xi \right)}{m^2 \xi
   }\,.
\end{equation}
In Fig. \ref{Case2Jordan}, we present the evolution of the Hubble parameter and the constrained scalar.

\subsubsection*{\color{Black}{$\diamond$} \textbf{{\textcolor{Black}{Case 3: Exponential expansion}}}}
Let us now consider the case when the Hubble parameter is constant. This corresponds the case  $\alpha=\beta$, which gives
\begin{equation}\label{H_case3}
   H(t)= \frac{ m}{2 \sqrt{3\xi}}
\end{equation}
In this case, the general solution for the constrained scalar is given by
\begin{equation}
    \Theta = \Mpl^{2}\left(
-\frac{4 \Lambda}{m^{2}}+\frac{1}{\xi}
+c_{\Theta}\,{\mathrm e}^{\frac{ m\, t}{2 \sqrt{3\xi}}}\right)\,.
\end{equation}
If we further assume $\dot{\Theta}=0$, we find
\begin{equation}\label{th_case3}
    \Theta(t)=\Mpl^{2} \left(\frac{1}{\xi}-\frac{4\Lambda}{m^2} \right)\,.
\end{equation}
The positivity of $\Theta$ is guaranteed if $\Lambda<0$, or
for $\xi<m^2/(4\Lambda)$ if $\Lambda>0$.

As we have previously mentioned, the solution for the Hubble parameter has a symmetry,
\begin{equation}
    \mu^2=-m^2>0\quad\text{and }\quad \zeta=-\xi>0\,,
\end{equation}
in which case we have 
\begin{equation}
     H=\frac{\mu}{2\sqrt{3\zeta}}\,,
\end{equation}
and the general solution for the scalar field is given by
\begin{equation}
    \Theta = \Mpl^2
    \left(
\frac{4 \Lambda}{\mu^{2}}-\frac{1}{\zeta}
+c_{\Theta}\,{\mathrm e}^{\frac{ \mu\, t}{2 \sqrt{3\zeta}}}\right)\,.
\label{Thetasol}
\end{equation}
In this case, one must have $\Lambda>0$ and
$\zeta>{\mu^2}/{(4\Lambda)}$.

\subsubsection*{\color{Black}{$\diamond$} \textbf{{\textcolor{Black}{Case 4: The general case}}}}
\begin{figure}[h!]
    \centering

    \begin{minipage}{0.45\textwidth}
        \centering
        \includegraphics[width=\linewidth]{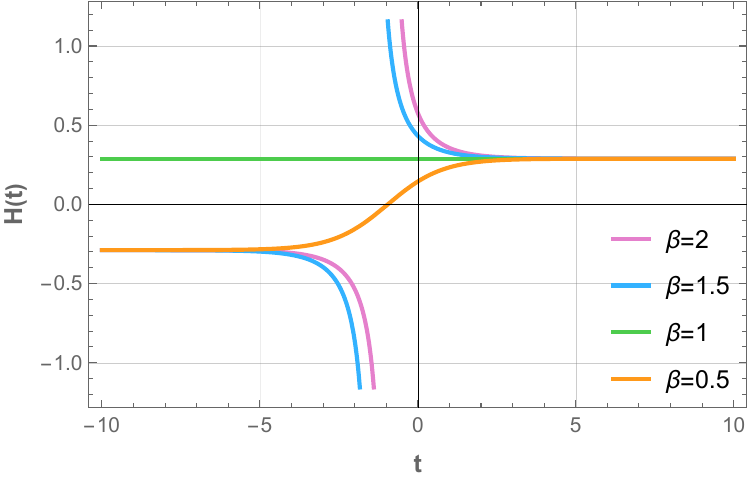}
    \end{minipage}
    \hfill
    \begin{minipage}{0.44\textwidth}
        \centering
        \includegraphics[width=\linewidth]{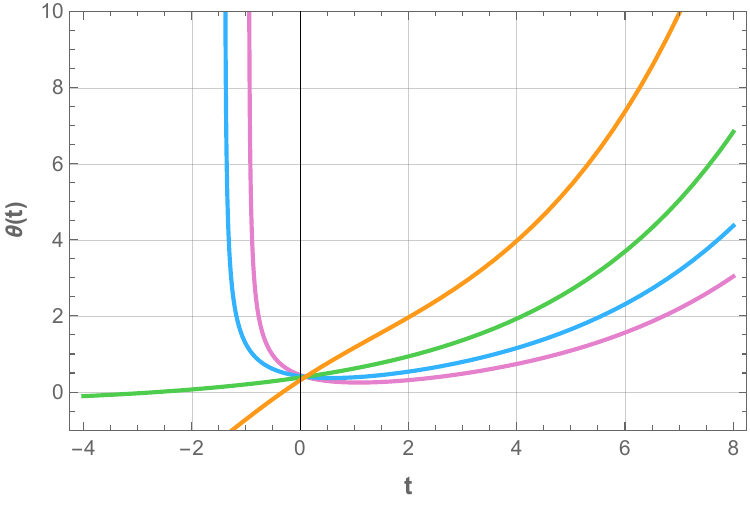}
    \end{minipage}

    \caption{{Solutions in Case 4 for the Hubble parameter (left) and the constrained scalar (right).  
    The legend given in the left panel applies also to the right panel. 
    The parameters are $m=1$, $\xi=1$, $\Lambda=1/3$,
    $\alpha=1$ and $\Theta(0.1)=0.4$ in the reduced Planck units $\Mpl=1$, for various values of $\beta$.
} }
    \label{HThetaJordan}
\end{figure}

The general case depends on the integration constants $\alpha$ and $\beta$,
\begin{equation}
    H(t)=\frac{\sqrt{3}m}{6\sqrt{\xi}}\frac{\alpha\sinh\gamma+\beta\cosh\gamma}{\alpha\cosh\gamma+\beta\sinh\gamma}\,,\quad\text{where}\quad\gamma=\frac{mt}{\sqrt{3\xi}}\,.
\end{equation}
Here, we assume both $\alpha$ and $\beta$ are positive for simplicity.
For $\beta>\alpha$, the solution behaves as Case 3, whereas for $\alpha\gg\beta$, it is close to the Case 1 with small corrections. Thus, the asymptotics of the scalar field will be similar to the previous cases.  In Fig. \ref{HThetaJordan}, we present the behavior of the scalar field, evaluated numerically, and the corresponding behavior of the Hubble parameter. 

\subsection{\textbf{{\textcolor{Black}{The massless theory}}}}
So far we have always assumed that the scalar must be massive, $m\neq0$.
However, it is also possible to realize the case when its mass is zero. 
For completeness, we will consider this case in this subsection. In particular, the vanishing of the mass of the constrained scalar modifies the constraint such that the Ricci scalar vanishes,
\begin{equation}
    R=0\,,
\end{equation}
hence the Hubble parameter mimics that of a radiation dominated universe, 
\begin{equation}
    \dot{H}=-2H\qquad\to\qquad H=\frac{1}{2t}\,.
\end{equation}
In this case, \eqref{eomTh} can be easily solved to give
\begin{equation}
    \Theta(t)=\frac{c_{\Theta}}{\sqrt{t}}+\frac{\Mpl^2(4t^2\Lambda-15)}{15\xi}
\end{equation}
This function can always be positive by a suitable choice of coupling. 

Although this recovery of the radiation dominated behavior in the massless limit is interesting in its own right, in the reminder of this work, we will focus on the more general case that involves the mass term. 

\section{Degrees of freedom in the Jordan frame}
\label{sec:dof}

Models with non-minimal coupling are particularly interesting from the point of view of the count in the degrees of freedom. If we would consider the scalar field without it, then, the number of degrees of freedom is two, corresponding to the two tensor modes that arise from the Einstein gravity. The non-minimal term, however, can change this, and give rise to a constrained scalar. An interesting case where this happens is in Proca theory with non-minimal coupling \cite{DeFelice:2025ykh}.

In this section, we will explore the type and behavior of the modes for our model, in the Jordan frame. 
In order to see the nature of these degrees of freedom, let us study the perturbations around the homogeneous and isotropic universe. We will assume that the background equations of motion are always satisfied. Then, we perturb around this background, 
\begin{equation}
    g_{\mu\nu}=g_{\mu\nu}^{(0)}+\delta g_{\mu\nu}\qquad\text{and} \qquad \sigma=\sigma(t)+\delta\sigma,
\end{equation}
where
\begin{equation}
    \begin{split}
        \delta g_{00}&=-2\phi\,,\\
        \delta g_{0i}&=a(t)\left(S_i+B_{,i}\right)\,,\\
        \delta g_{ij}&=a^2(t)\left(2\psi \delta_{ij}+2E_{,ij}+F_{i,j}+F_{j,i}+h_{ij}^T\right)\,,\\
    \end{split}
\end{equation}
with the conditions,
\begin{equation}
    S_{i}{}^{,i}=0\,,\quad F_i{}^{,i}=0\,,\quad h^T_{ij}{}^{,j}=\delta^{ij}h^T_{ij}=0\,.
\end{equation}
We will work in the gauge,
\begin{equation}
    E=B=F_i=0\,,
\end{equation}
for which the modes coincide with the gauge-invariant variables. It is easy to find that for the above decomposition, the scalar, vector, and tensor modes decouple, which allows us to study them separately.

For the scalar modes, we first note that $\phi$ does not propagate. Accordingly, we solve the constraint equation and substitute it back to the action. 
Then, after several integrations by parts, if we perform a substitution,
\begin{equation}
    \delta\sigma= \delta{\sigma}_f+\frac{\psi \dot{\sigma}}{H}\,,
\end{equation}
we find that $\psi$ is constrained. By solving the constraint and substituting back to the action, we find a Lagrangian density for the single scalar mode $\delta\sigma_f$, which is in the momentum space given by
\begin{equation}
\mathcal{\tilde{L}}_S=A(t)a^3\left(\delta\dot{{\sigma}}_{f\Vec{k}}\delta\dot{{\sigma}}_{f(-\Vec{k})}-\left(\frac{k^2}{a^2}+B(t)\right)\delta{\sigma}_{f\Vec{k}}\delta{\sigma}_{f(-\Vec{k})}\right),
\end{equation}
where $A(t)$ and $B(t)$ are coefficients that depend on time. 
In the terminology of cosmological perturbation theory, $\delta\sigma_f$ corresponds to the scalar field perturbation on flat slices.
We immediately see that the speed of the propagation for the constrained scalar is unity,
$c_{\sigma}^2=1$. 

For the perturbations to be healthy, we require the non-ghost condition, $A(t)>0$. Explicitly, it is given by
\begin{equation}
    A(t)=\frac{3\xi^{2} \sigma^{2} \left(\xi  \sigma^{2}+\Mpl^2\right)}{{\left(\xi\sigma\dot{\sigma}/H+\left(\xi  \sigma^{2}+\Mpl^2\right)\right)}^{2}}>0\,.
\end{equation}
As clearly seen from the above, if we assume $\sigma$ to be real, $\xi>0$ is a sufficient condition for $\delta{\sigma_f}$ to be a healthy scalar mode.
Note also that, if one allows $\sigma^2=\Theta$ to become negative, the non-ghost condition becomes non-trivial, $\Theta(\xi\Theta+\Mpl^2)>0$. 

The $B$ coefficient is given by
\begin{equation}
   \begin{split}
       B(t)&=\frac{1}{6\xi^2\sigma^2(\xi\sigma^2+\Mpl^2)\left( \xi  \sigma^{2}+\Mpl^{2}+\xi  \sigma \dot{\sigma}/H\right)}\Bigl[12 \xi  \,H^{2} \left(\xi  \sigma^{2}+\Mpl^{2}\right)^{2} \left(3 \xi  \sigma^{2}+\Mpl^{2}\right)\\
       &+12 H \sigma\dot\sigma\xi^{2} \left(\xi  \sigma^{2}+\Mpl^{2}\right) \left(3 \xi  \sigma^{2}+\Mpl^{2}\right)-(\xi\sigma^2)^2\left(5 m^{2} \xi \sigma^{2}+66 \xi^{2} \dot{\sigma}^{2}+11 \Mpl^{2}m^2\right)\\
      &+(\xi \sigma \dot{\sigma}/H)\left(-\xi  \sigma^{2}+\Mpl^{2}\right) \left(m^{2} \xi  \sigma^{2}-6 \xi^{2} \dot{\sigma}^{2}+\Mpl^{2}m^2\right) +2 m^{2} (\xi\sigma\dot\sigma/H)^2 \left(2 \xi  \sigma^{2}+\Mpl^{2}\right)\\
       &
       +\Mpl^2\left(36\xi^{3} \dot{\sigma}^{2}\sigma^2+ \Mpl^{2}(7m^2 \xi\sigma^2+6 \xi^{2} \dot{\sigma}^{2})+\Mpl^{4} m^{2}\right) \Bigr].
   \end{split}
\end{equation}
At late times, all the solutions tend to the constant Hubble parameter given by (\ref{H_case3}). 
Also, except for the spacial case when the coefficient $c_\Theta$ in (\ref{Thetasol}) vanishes, $\Theta=\sigma^2$ grows exponentially as $\Theta\propto e^{Ht}$.
By substituting this into the previous expression, we find in the late time limit $t\to\infty$,
\begin{equation}
    B(t)\to-\frac{7}{4}H^2=-\frac{7m^2}{48\xi}\,.
\end{equation}
Thus the perturbation is tachyonic in the late time limit. This is reasonable since the background $\sigma$ is exponentially growing. 
In passing, we note that this also indicates the instability of the special constant solution $\Theta=const.$ for $c_\Theta=0$ in (\ref{Thetasol}). 
This tachyonic instability could lead to an undesirable consequences if it would persist forever. 
However, we mention again that the current model is never meant to be realistic. 
At least, we have to incorporate a graceful exit from the de Sitter phase to make the model realistic.
The incorporation of a graceful exit is expected to tame the instability.

As clear from the above, while the scalar field is constrained at the background level, its fluctuation has a propagating degree of freedom.
This implies this degree of freedom will naturally arise if we treat $\sigma$ quantum field theoretically. 
The origin of the propagating scalar degree of freedom is its non-minimal coupling to the Ricci scalar curvature which introduces coupling between 
$\delta\sigma$ and the scalar-type gravitational (or gravitational potential) perturbations.
This is in essence similar to the temporal component of the vector field, which is constrained, but if non-minimally coupled gives rise to extra propagating mode due to its coupling with the gravitational perturbations \cite{DeFelice:2025ykh}. 
We also mention that our theory is similar to the freezing gravity proposed in \cite{Yao:2025wlx}, which appeared after our work.

We note that $A(t)$ vanishes if $\sigma^2=0$. This indicates a potential strong coupling for the scalar mode in the limit $\sigma\to 0$. In fact, in Case 1 discussed in the previous section, if we consider the initial singularity to be at $\sigma=0$, the scalar mode will be strongly coupled in this limit. Thus the non-linear terms may become important, signalling the breakdown of perturbation theory. 
This suggests a scenario in which the Universe started from a strongly coupled regime, and has then undergone the super-Hubble expansion, followed by an exponential expansion.
Again, one should nevertheless keep in mind that the simplest case we focus on in this work is far from being realistic. We would need more ingredients or more sophisticated models to make the constrained scalar framework viable as a theory of the early universe.

On the other hand, in Case 2 discussed in the previous section, $\sigma$ never vanishes. Moreover, $A(t)$ remains finite in the limit $t\to0$ as well. As for Case 4 (ie, the general case), the similar result holds if $\beta>\alpha$, while the scalar will becomes strongly coupled if $\beta<\alpha$. 
The case $\alpha=\beta$ is identical to Case 3 (ie, eternal de Sitter), and $\sigma=0$ may occur during the course of evolution, depending on the choice of the model parameters.

For the vector modes, one can easily see that they do not propagate. On the other hand, the tensor modes propagate as usually the case for Brans-Dicke type scalar-tensor theories.
Its Lagrangian can be easily obtained as
\begin{equation}
    \mathcal{L}_T=\frac{a(\xi\sigma^2+\Mpl^2)}{8}\left(a^2\dot{h}_{ij}^T\dot{h}_{ij}^T-k^2h_{ij}^Th_{ij}^T\right)\,.
\end{equation}
Apparently, in the case one allows negative $\sigma^2$ ($=\Theta$), the non-ghost condition is the same as the one for the scalar mode, i.e., $\xi\Theta+\Mpl^2>0$

To summarize, our theory describes one scalar and two tensor degrees of freedom, with the no-ghost condition being always satisfied if $\xi>0$ and $\sigma^2>0$, or as long as $\xi\sigma^2+\Mpl^2>0$ even if one allows negative $\sigma^2$. 
As we will see in the next section, another confirmation of this is given in the Einstein frame. 

\section{The Einstein frame}
In the previous section, we studied the model of the constrained scalar with non-minimal coupling in the Jordan frame. However, it is often more convenient to translate its evolution to the Einstein frame, where the Ricci scalar is minimally coupled to the scalar field. In this section, we will explore our theory in this frame, and connect the solutions between this and the Jordan frame as well.

The metric in the Einstein frame $\tilde{g}_{\mu\nu}$ is obtained by the conformal transformation,
\begin{equation}
g_{\mu\nu}~\to~   \tilde{g}_{\mu\nu}=F g_{\mu\nu}\,,
\end{equation}
where $F$ is given by 
\begin{equation}
    F=1+\frac{\xi}{\Mpl^2}\sigma^2\,.
\end{equation}
Accordingly, the Ricci scalar in the Jordan-frame is expressed in terms of the Einstein-frame quantities as
\begin{equation}
    R=F\tilde{R}+3\tilde{\nabla}_{\mu}\tilde{\nabla}^{\mu}F-\frac{9}{2}\frac{1}{F}\tilde{\nabla}_{\mu}F\tilde{\nabla}^{\mu}F\,,
\end{equation}
where $\tilde{R}$ and $\tilde{\nabla}$ are the Ricci tensor and the covariant derivative, respectively, in the Einstein frame. 
Plugging the above into the action, we obtain the Einstein-frame action,
\begin{equation}
    S=\int d^4x\sqrt{-\tilde{g}}\left[\frac{\Mpl^2}{2}\tilde{R}-\frac{3}{4}\frac{\Mpl^2}{F^2}\tilde{\nabla}_{\mu}F\tilde{\nabla}^{\mu}F-\frac{\Mpl^2}{F^2}\Lambda-\frac{m^2}{2}\frac{\Mpl^2}{\xi F^2}(F-1)\right].
\end{equation}
 By further defining
\begin{equation}
    \phi=\sqrt{\frac{3}{2}}\Mpl\ln F, 
\end{equation}
the above action becomes
\begin{equation}\label{EinsteinFrameAction}
    S=\int d^4x\sqrt{-\tilde{g}}\left[\frac{\Mpl^2}{2}\tilde{R}-\frac{1}{2}\tilde{\nabla}_{\mu}\phi\tilde{\nabla}^{\mu}\phi-V(\phi)\right],
\end{equation}
where
\begin{equation}
    V(\phi)=\Mpl^2e^{-2\sqrt{\frac{2}{3}}\frac{\phi}{\Mpl}}\left[\Lambda-\frac{m^2}{2\xi}+\frac{m^2}{2\xi}e^{\sqrt{\frac{2}{3}}\frac{\phi}{\Mpl}}\right].
\end{equation}
Therefore, in the Einstein frame, our theory, initially defined with a constrained scalar with non-minimal coupling, has become a theory of Einstein-Hilbert action and a scalar field with a potential. Curiously, we can right away notice that the dynamics of this frame will depend on the cosmological constant $\Lambda$ of the Jordan frame, although the dynamics of space-time in the Jordan frame were independent of it. 
We also note that the theory in the Einstein frame is well-defined as long as $F>0$, which is equivalent to the non-ghost condition in the Jordan frame if one allows negative values of $\Theta=\sigma^2$.

Let us now explore the solutions in the Einstein frame. 
After that, we discuss the relation between the solutions in the two frames. In the next three subsections, all quantities will be those defined in the Einstein frame unless otherwise stated.

\subsection{The accelerating solutions} 

As a first step, we will search for the background equations of motion corresponding to the action in the Einstein frame (\ref{EinsteinFrameAction}).
For this, we will assume again an FLRW Universe, 
\begin{align}
   &ds_E^2=-N^2dt_E^2+a(t_E)^2\delta_{ij}dx^i_Edx^j_E\,, 
   \nonumber\\
   &\phi=\phi(t_E)\,,
\end{align}
where the subscript $E$ means the Einstein frame.
The lapse function $N$ is inserted for convenience, but will be set to unity after the field equations have been derived. 
For notational simplicity, in this and the next two subsections, 
we omit the subscript $E$, and the dot ($\dot{~}$) denotes the derivative with respect to the cosmic time in the Einstein frame. 
Thus we have
\begin{equation}
    t=t_E\quad \text{and}\quad \dot{X}=\frac{dX}{dt_E}\,.
\end{equation}

By varying the action (\ref{EinsteinFrameAction}) with respect to the lapse $N$, and setting $N=1$ afterwards, we find the Hamiltonian constraint, which we rewrite as
\begin{equation}
    \Mpl^{2} \left(-2 \Lambda \xi +m^{2}\right) {\mathrm e}^{-\frac{2 \sqrt{6}\, \phi \left(t \right)}{3 \Mpl}}-\Mpl^{2} {\mathrm e}^{-\frac{\sqrt{6}\, \phi \left(t \right)}{3 \Mpl}} m^{2}- \xi  \left(-6 \Mpl^{2} H^{2}+\dot{\phi}^{2}\right)=0\,.
    \label{Hconst}
\end{equation}
The acceleration equation can be obtained by varying the action with respect to the scale factor $a$,
\begin{equation}
    \Mpl^{2} \left(-2 \Lambda \xi +m^{2}\right) {\mathrm e}^{-\frac{2 \sqrt{6}\, \phi \left(t \right)}{3 \Mpl}}-\Mpl^{2} {\mathrm e}^{-\frac{\sqrt{6}\, \phi \left(t \right)}{3 \Mpl}} m^{2}+4\xi \left(\frac{\dot{\phi}^{2}}{4}+\Mpl^{2} \left(\frac{3 H^{2}}{2}+\dot{H}\right)\right)=0\,.
    \label{acceq}
\end{equation}
In addition, by varying the action with respect to the scalar field $\phi$, we find
\begin{equation}
    \sqrt{6}\Mpl\, \left(-2 \Lambda \xi +m^{2}\right) {\mathrm e}^{-\frac{2 \sqrt{6}\, \phi \left(t \right)}{3 \Mpl}}-\frac{\Mpl \sqrt{6}\, {\mathrm e}^{-\frac{\sqrt{6}\, \phi \left(t \right)}{3 \Mpl}} m^{2}}{2}+9 \xi\left(\dot{\phi} H +\frac{\ddot{\phi}}{3}\right) =0\,.
   \label{phieq}
\end{equation}
As well known, \eqref{acceq} is automatically satisfied provided that \eqref{Hconst} and \eqref{phieq} are both satisfied.
It is worth mentioning that, by solving \eqref{Hconst} for $H^2$, and substituting it into \eqref{acceq}, we find 
\begin{equation}
    2\Mpl^2\dot{H}+\dot{\phi}^2=0\,. 
\end{equation}

Note that in the special case, 
\begin{equation}
    2\Lambda \xi = m^2,
\end{equation}
the potential corresponds to the power law solution \cite{Lucchin:1984yf}, with an analytical solution,
\begin{equation}
    H(t)=\frac{3}{t}\,,\quad \phi=\sqrt{6}\Mpl\ln(t/t_*)\,,
    %\quad m^2=48\xi\,,
\end{equation}
where $t_*$ is a constant determined by the value of $\xi/m^2$.

\subsection{The slow-roll conditions} 
Let us now consider the slow-roll conditions, 
\begin{equation}\label{slowrollconditions}
    |\dot{\phi}^2|\ll V(\phi) \qquad \text{and}\qquad |\ddot{\phi}|\ll 3H\dot{\phi}\,.
\end{equation}
In this case, the Hamiltonian constraint and the scalar field equation can be approximated as
\begin{equation}
    \begin{split}
        H(t)^2 =&
\frac{1}{6 \xi}{\mathrm e}^{-\frac{\sqrt{6}\, \phi}{3 \Mpl}} \left((2 \Lambda \,\xi -m^{2}) {\mathrm e}^{-\frac{\sqrt{6}\, \phi}{3 \Mpl}}+m^{2}\right)\,,\\
    \dot{\phi}(t)=& 
\frac{\sqrt{6}\Mpl}{18 H \xi} \,{\mathrm e}^{-\frac{\sqrt{6}\, \phi}{3 \Mpl}} \left(2(2 \Lambda \, \xi - m^{2} ){\mathrm e}^{-\frac{\sqrt{6}\, \phi}{3 \Mpl}}+m^{2}\right)\,.
    \end{split}
\end{equation}
Let us assume $\xi>0$. Then, the first equation imposes the condition, 
\begin{equation}\label{conditionRealH}
    (2 \Lambda \,\xi -m^{2}) {\mathrm e}^{-\frac{\sqrt{6}\, \phi}{3 \Mpl}}+m^{2}>0\,.
\end{equation}
This is satisfied if $2\Lambda\xi\geq m^2$. 
In addition, by writing the slow-roll conditions in terms of the potential and its derivatives, we find the following conditions: 
\begin{equation}
\frac{2 \left(m^{2} {\mathrm e}^{\frac{\sqrt{6}\, \phi}{3 \Mpl}}+4 \Lambda\xi -2 m^{2}\right)^{2}}{9\left(m^{2} {\mathrm e}^{\frac{\sqrt{6}\, \phi}{3 \Mpl}}+2 \Lambda\xi -m^{2}\right)^{2}}\ll1 \qquad \text{and}\qquad \frac{2}{9}\left|\frac{ m^{2} {\mathrm e}^{\frac{\sqrt{6}\, \phi}{3 \Mpl}}+8 \Lambda\xi-4 m^{2}}{m^{2} {\mathrm e}^{\frac{\sqrt{6}\, \phi}{3 \Mpl}}+2 \Lambda\xi -m^{2}}\right|\ll1\,.
\end{equation}
In the limit $\phi\gg\Mpl$, the exponential factor dominates, and both conditions reduce to
\begin{equation}
    \frac{2}{9}\ll 1\,.
\end{equation}
Thus the slow-roll conditions are marginally satisfied. In fact, as we have seen at the end of the previous subsection, the approximate solution corresponds to the power law solution $a\propto t^p$ with $p=3$, which becomes exact in the case $2\Lambda\xi=m^2$.
Note, however, inflation is eternal in this model. One would need another field or matter to terminate inflation.

\subsection{Numerical solutions}

Knowing the slow-roll conditions gives us an intuition of under which conditions can we obtain an accelerating Universe. Let us now show some of the solutions that correspond to it, for particular choices of parameters. 
Namely, we again employ the reduced Planck units $\Mpl=1$, and choose 
\begin{equation}
    m=1\qquad \text{and}\qquad \xi=1\,.
\end{equation}
Then we solve the equations,
\begin{equation}
 \begin{split}
        \dot{H}&=-\frac{\dot\phi^2}{2\Mpl^2}\\
        \Ddot{\phi}&=\frac{\left(4 \Lambda \Mpl \,{\mathrm e}^{-\frac{2 \sqrt{6}\, \phi \left(t \right)}{3 \Mpl}} \xi -2 \Mpl \,{\mathrm e}^{-\frac{2 \sqrt{6}\, \phi \left(t \right)}{3 \Mpl}} m^{2}+\Mpl \,{\mathrm e}^{-\frac{\sqrt{6}\, \phi \left(t \right)}{3 \Mpl}} m^{2}-3 \sqrt{6}\, H \dot{\phi}\xi \right) \sqrt{6}}{6 \xi}\,,
 \end{split}
\end{equation}
with the initial conditions for $H$, $\dot{\phi}$ and $\phi$ such that they satisfy the Hamiltonian constraint.

\begin{figure}[h]
    \centering
    \begin{minipage}{0.3\textwidth}
        \centering
        \includegraphics[width=\linewidth]{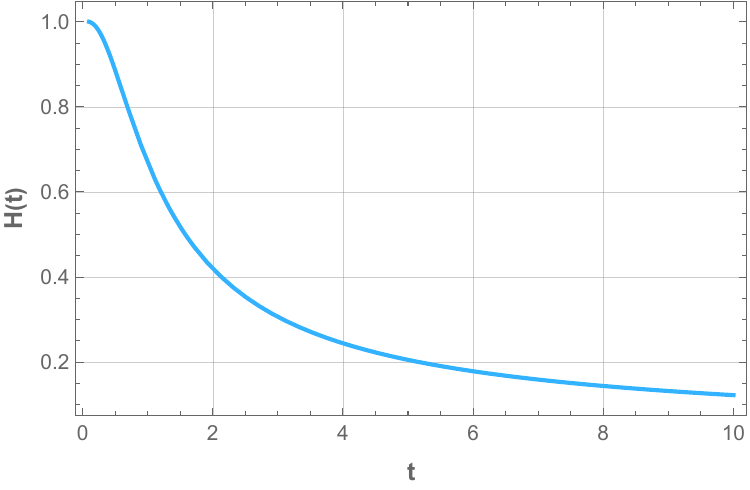}
    \end{minipage}
    \hfill
    \begin{minipage}{0.3\textwidth}
        \centering
        \includegraphics[width=\linewidth]{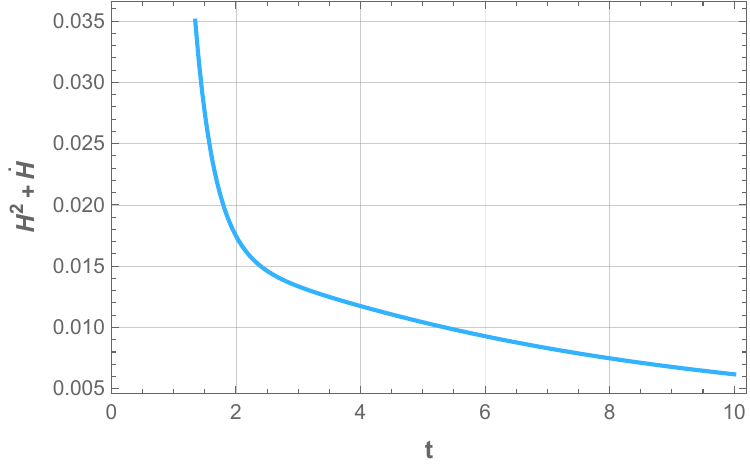}
    \end{minipage}
     \hfill
    \begin{minipage}{0.3\textwidth}
        \centering
        \includegraphics[width=\linewidth]{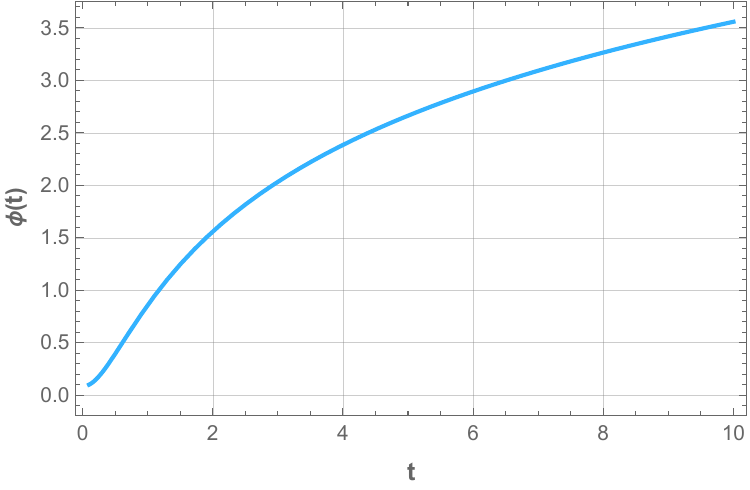}
    \end{minipage}
    \vspace{5mm}
    
    \begin{minipage}{0.3\textwidth}
        \centering
        \includegraphics[width=\linewidth]{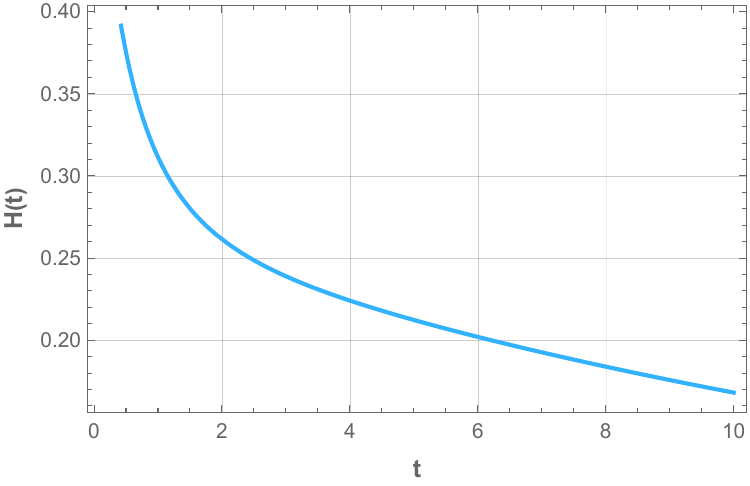}
    \end{minipage}
    \hfill
    \begin{minipage}{0.3\textwidth}
        \centering
        \includegraphics[width=\linewidth]{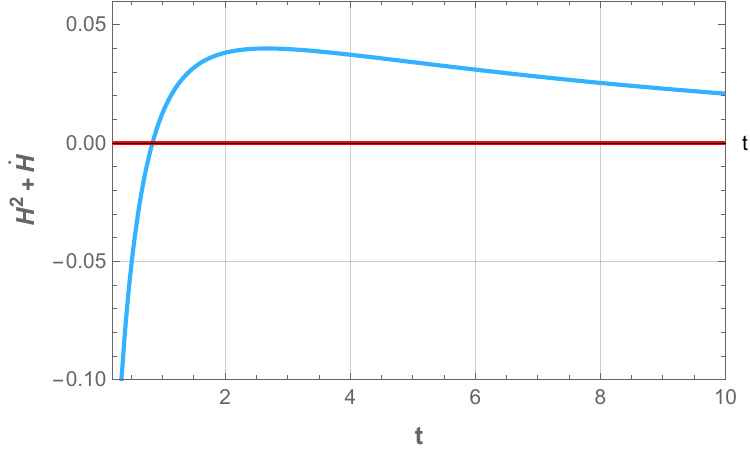}
    \end{minipage}
    \hfill
    \begin{minipage}{0.3\textwidth}
        \centering
        \includegraphics[width=\linewidth]{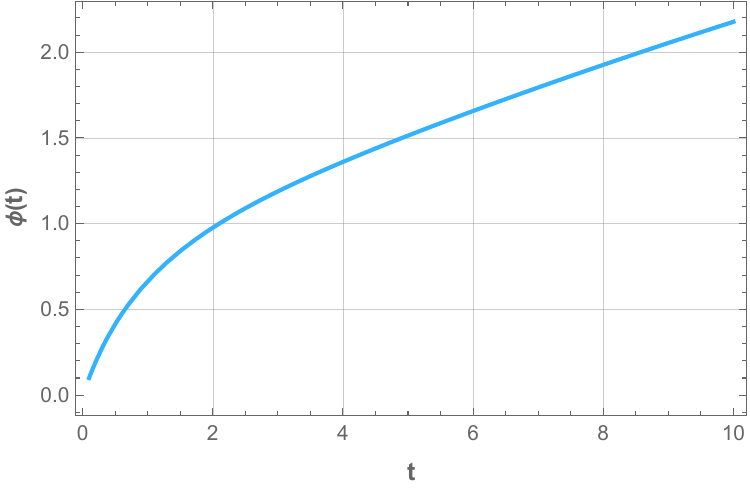}
    \end{minipage}
    \caption{The evolution of $H$, $\ddot{a}/a=H^2+\dot H$, and $\phi$ for two different sets of initial conditions. The upper panel is for the case when the universe is always in an accelerated expansion phase, while the lower panel is for the case when the universe begins in a decelerated phase and evolves to an accelerated phase.}
    \label{fig:threeinrow1}
\end{figure}

Specifically, we consider two sets of initial conditions. First one is 
\begin{equation}
    \phi(0.1)=0.1\,,\quad \phi'(0.1)=0.1\,,\quad H(0.1)=1\,,
\end{equation}
and the cosmological constant is determined by requiring the constraint to be satisfied, giving $\Lambda = 3.43665$. 
The resulting evolutionary behaviors of the Hubble parameter $H$, the acceleration $\ddot{a}/a=H^2+\dot H$, the scalar field $\phi$ are presented in the upper panel of Fig.~\ref{fig:threeinrow1}.
The second one is
\begin{equation}
    \phi(0.1)=0.1\,,\quad \phi'(0.1)=1\,,\quad H(0.1)=0.5,
\end{equation}
which gives $\Lambda=0.25$. 
The results are presented in the lower panel of Fig.~\ref{fig:threeinrow1}.
We see that this model allows for a transition from decelerating to an accelerating Universe, depending on the model parameters. 

Overall, we can see that our theory is very curious: 
In the Jordan frame, the space-time evolves completely independently of the cosmological constant. 
In the Einstein frame, on the contrary, the cosmological constant enters directly into the scalar field potential, hence it affects the evolution crucially. 
Moreover, the Einstein frame allows a transition from a decelerated to an accelerated phase, similar to Case 2 of the Jordan frame.
However, it is not clear at all which solution in the Einstein frame corresponds to which solution in the Jordan frame. 
To see this, let us consider the correspondence between the solutions of each frame. 

\subsection{Connection between the Jordan and Einstein frame}

To compare the solutions of each frame, we recover the subscript $E$ to quantities in the Einstein frame.
The metric in the Einstein frame is related to that in the Jordan frame as
\begin{equation}
    ds_E^2=Fds^2=Fg_{\mu\nu}dx^{\mu}dx^{\nu}=F(-dt^2+a^2\delta_{ij}dx^idx^j),
\end{equation}
where $ds^2$ is the Jordan-frame metric, $t$ is the cosmic time in the Jordan frame, and
\begin{equation}
    F=1+\frac{\xi}{\Mpl}\Theta(t). 
\end{equation}
Thus comparing the above with the expression in the Einstein frame, 
\begin{equation}
    ds^2_E=-dt_E^2+a^2_E\delta_{ij}dx^i_Edx^j_E\,, 
\end{equation}
we have \cite{Domenech:2016yxd}
\begin{equation}
    dt_E=\sqrt{F}dt\,,\quad a_E=\sqrt{F}a\,,\quad x_E=x\,.
\end{equation} 
The Hubble parameter in the Einstein frame is related to the one in the Jordan frame by
\begin{equation}
    H_E=\frac{1}{a_E}\frac{d a_E}{dt_E}= \frac{\dot{F}}{2F^{3/2}}+\frac{H}{\sqrt{F}},
\end{equation}
where the dot denotes a time derivative $d/dt$ in the Jordan frame, and $H$ is the Hubble parameter in the Jordan frame. 
In addition, the scalar field $\phi$ is directly related to the scalar field of the Jordan frame via:
\begin{equation}
        \phi=\sqrt{\frac{3}{2}}\Mpl\ln F\,.
\end{equation}

Let us now find the solutions in the Einstein frame which correspond to the solutions for Cases 1 and 2 in the Jordan frame. For simplicity, we set  
\begin{equation}
    m=1\,,\quad\xi=1\,,\quad \Lambda=1/3\,.
\end{equation}
\subsubsection*{\color{Black}{$\diamond$} \textbf{{\textcolor{Black}{Case 1: Super-accelerated to exponential expansion}}  } }
\begin{figure}[h!]
    \centering

    \begin{minipage}{0.45\textwidth}
        \centering
        \includegraphics[width=\linewidth]{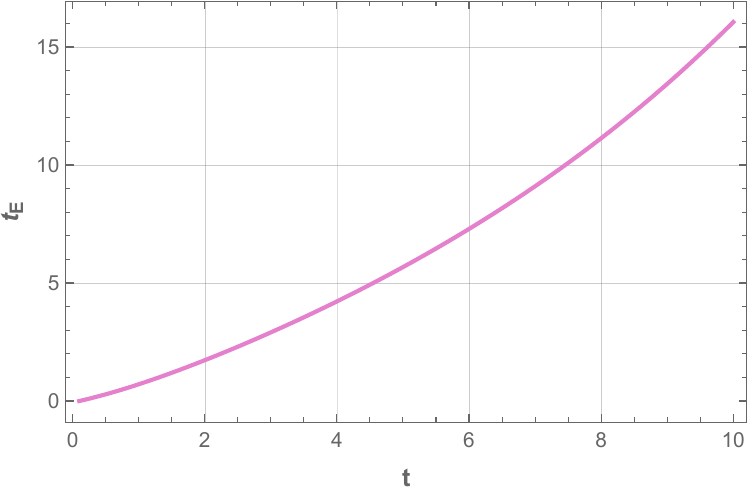}
    \end{minipage}
    \hfill
    \begin{minipage}{0.45\textwidth}
        \centering
        \includegraphics[width=\linewidth]{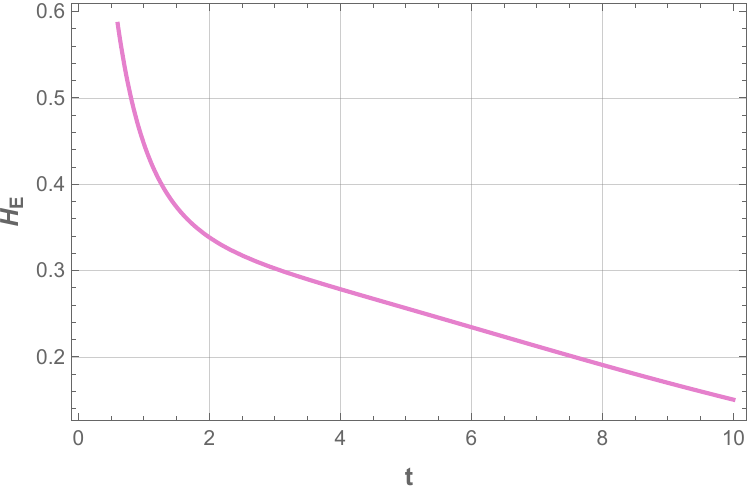}
    \end{minipage}

     \centering

    \begin{minipage}{0.45\textwidth}
        \centering
        \includegraphics[width=\linewidth]{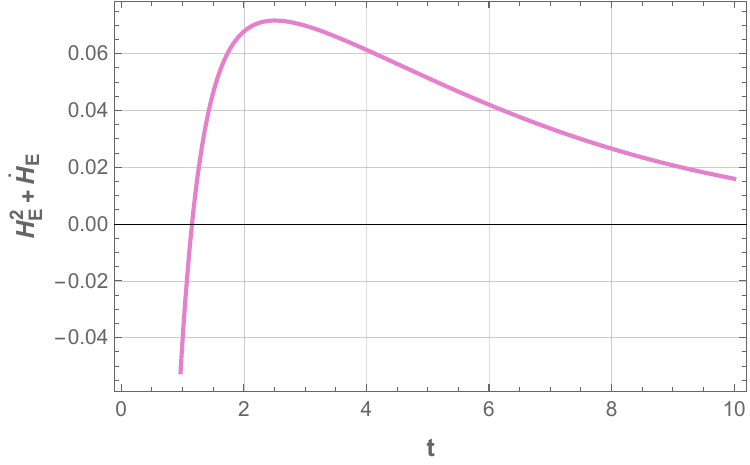}
    \end{minipage}
    \hfill
    \begin{minipage}{0.45\textwidth}
        \centering
        \includegraphics[width=\linewidth]{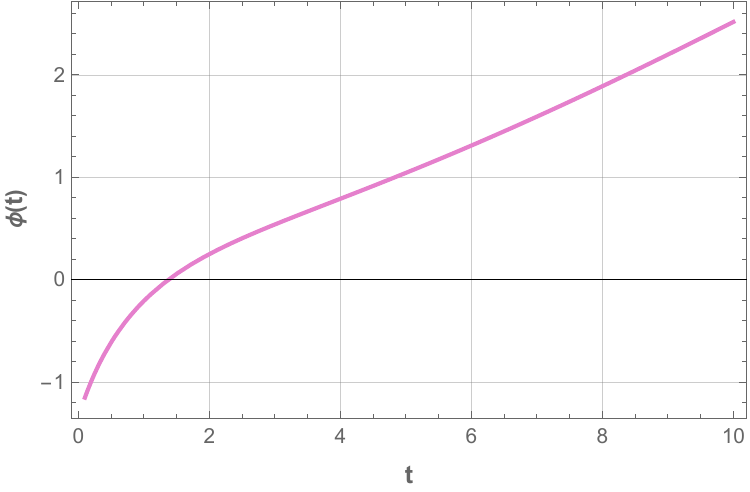}
    \end{minipage}

    \caption{
    { The relation between the Einstein frame, and the solution of the super-accelerated to exponential expansion in the Jordan frame described by equations (\ref{JC1Hub}) and (\ref{Case1Sol}).}
    \\
    {\bf Upper panel:}{ The left figure shows the cosmic time in the Einstein frame as a function of the Jordan-frame time. The right figure shows the Hubble parameter in the Einstein frame.}  
\\
  {\bf Lower panel:}{ The left figure shows the acceleration $\ddot{a}_E/a_E$ and the right figure the scalar field as functions of the Jordan-frame time.
  In the Einstein frame, it appears to be decelerating, and then transitioning to an accelerating stage. 
} }
    \label{super-acc_E}
\end{figure}

This solution corresponds to the Fig.~\ref{Case1JThetaH}. 
To find the Einstein time, we find the Einstein time in terms of the Jordan one, by integrating over the square root of $F$ and taking into account that the Jordan time starts at $t_0$, the value at which $\Theta$ vanishes, so that the constrained scalar is real. 
For simplicity, we present the values of the Hubble parameter, the acceleration and the scalar field in terms of the Jordan time in Fig.~\ref{super-acc_E}. 

\subsubsection*{\color{Black}{$\diamond$} \textbf{{\textcolor{Black}{Case 2: Radiation dominated to exponential expansion}}  } }
\begin{figure}[h!]
    \centering

    \begin{minipage}{0.45\textwidth}
        \centering
        \includegraphics[width=\linewidth]{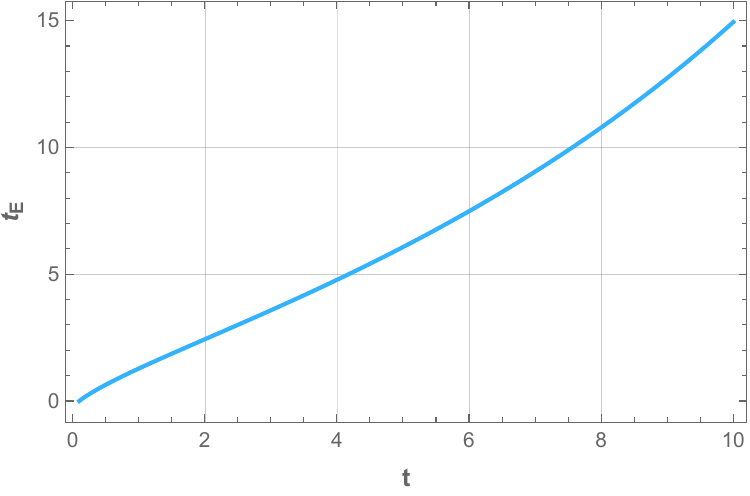}
    \end{minipage}
    \hfill
    \begin{minipage}{0.45\textwidth}
        \centering
        \includegraphics[width=\linewidth]{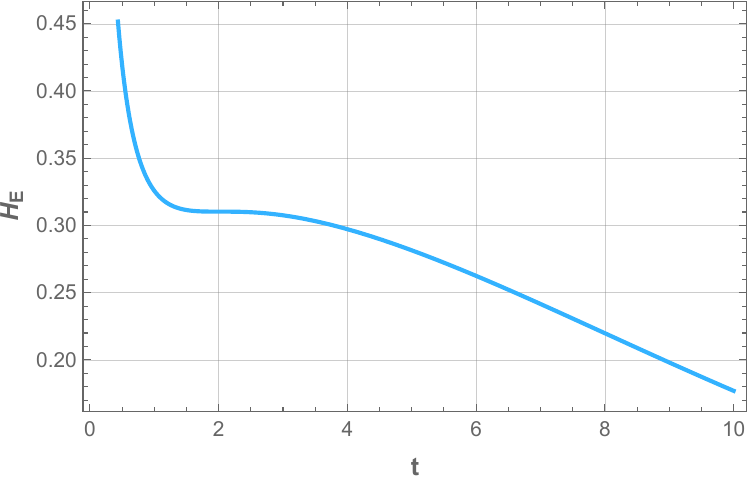}
    \end{minipage}

\centering

 \begin{minipage}{0.45\textwidth}
        \centering
        \includegraphics[width=\linewidth]{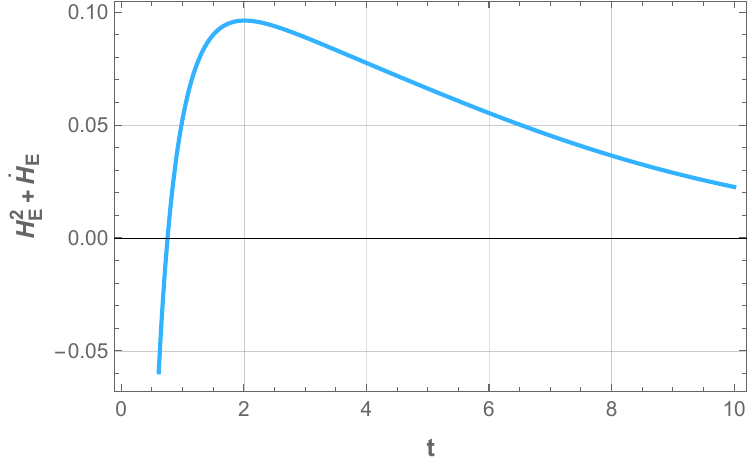}
    \end{minipage}
    \hfill
    \begin{minipage}{0.45\textwidth}
        \centering
        \includegraphics[width=\linewidth]{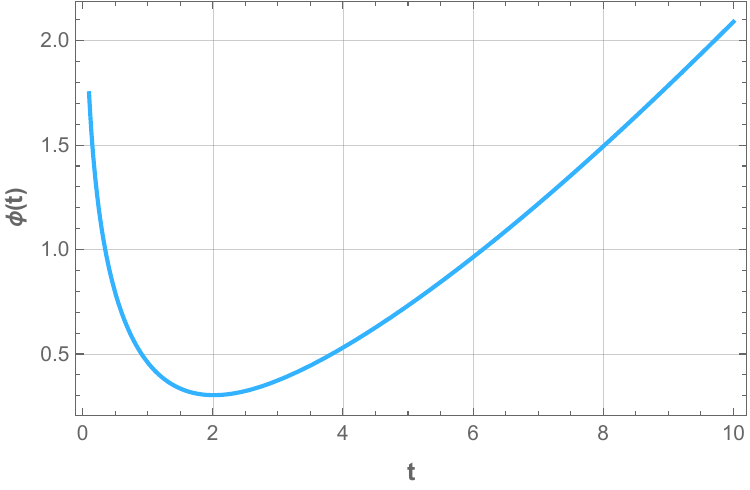}
    \end{minipage}

\caption{ { The relation between the Einstein frame, and the solution that describes the radiation dominated to exponential expansion found in the Jordan frame in (\ref{case2H}) and (\ref{solCase2}).}
\\
    {\bf Upper panel:}{ The figure on the left shows how times in the two frames are connected.  On the right, we plot the behavior of the Hubble parameter for this solution.   }  
    \\
  {\bf Lower panel:}{ The evolution of the scalar field (right) and the change in acceleration (left) as functions of the Jordan-frame time. In the Einstein frame, it is first decelerating, and then transitions to an accelerated stage.  } 
}\label{decc-accE}
\end{figure}

The solutions in this case are depicted in Figure \ref{Case2Jordan}, with the same initial conditions. By expressing the Einstein frame quantities in terms of them, we find the results that are depicted on the Figure \ref{decc-accE}.

\section{External matter}
In the previous sections, we have studied the theory of a non-minimal constrained scalar in the absence of matter. 

{We have seen that the solutions found in the previous sections behave in an very intriguing way -- although they make the Ricci scalar constant, they give rise to solutions which initially behave like radiation dominated Universe, which transitions to an accelerated expansion, or solutions that first expand via super-Hubble expansion, further then relaxing to an accelerated expansion -- in the absence of any matter, while at the same time being free of ghost.  While at this point such solutions do not describe a cosmological scenario that might describe our currently observed Universe,  they provide a good foundation to study the basic properties of our model of constrained scalar. To make a further step in analysing it, in this section, we will study the basic properties of this theoretical framework in the presence of matter.  }

Usually, the physical frame is defined as one where matter is minimally coupled to gravity. 
In our case, the Jordan frame shows very interesting behavior. 
So we will consider matter minimally coupled to gravity in the Jordan frame. 
However, in this case, we have two interesting choices -- we can couple the matter to the constrained scalar either minimally or non-minimally.
In this section, we will study these two possibilities. 
\subsection{The minimal coupling}
Let us first consider the case when the matter is minimally coupled to both the constrained scalar and gravity. In this case, the action given by
\begin{equation}
    \begin{split}
        S=\int d^4x\sqrt{-g}&\left[\frac{\Mpl^2}{2}\left(R-2\Lambda\right)-\frac{m^2}{2}\sigma^2+\frac{\xi}{2}R\sigma^2\right]+S_m\,.
    \end{split}
\end{equation}
For definiteness, we consider the case when $S_m$ is given by that of a scalar field $\chi$, 
\begin{equation}
    S_m=\int d^4x\sqrt{-g}\left(-\frac{1}{2}\partial_{\mu}\chi\partial^{\mu}\chi-V(\chi)\right)
\end{equation}
or a perfect fluid, 
\begin{equation}
     S_m=\int d^4x\sqrt{-g}\,p\,, 
\end{equation}
where $p$ is the pressure.

By varying the action with respect to the lapse, and setting it to unity afterwards, we obtain the constraint equation,
\begin{equation}\label{constraintMatter}
    3H^2(\Mpl^2+\xi\sigma^2)=\frac{1}{2}m^{2} \sigma^{2}-6 \sigma  H \dot{\sigma} \xi %+\Lambda \,\Mpl^{2}
    +\frac{1}{2}\dot{\chi}^{2}+ U\left(\chi \right)\,,
\end{equation}
where we have absorbed the cosmological constant to the potential $U$.
Note that this reduces to the standard Friedman equation when $\sigma=0$,
\begin{equation}
   {3\Mpl^2} H^2=\varepsilon_{\chi},\qquad\text{where}\qquad \varepsilon_{\chi}=\frac{1}{2}\dot\chi^2+U(\chi).
\end{equation}
We can also easily generalize \eqref{constraintMatter} to the case of a perfect fluid,
\begin{equation}
     3{\Mpl^2}H^2\left(1+\frac{\xi}{\Mpl^2}\Theta\right)=\varepsilon+ 
     \frac{1}{2}\left(m^2\Theta-6\xi H\dot\Theta\right)\,,
\end{equation}
where $\varepsilon$ is the energy density that includes the vacuum energy, and $\Theta=\sigma^2$ as before.

By varying the action with respect to the constrained scalar $\sigma$, we find
\begin{equation}
    \sigma  \left(-12 \xi  \,H^{2}-6 \xi  \dot{H} +m^{2}\right)=0\,,
\label{sigmaconstr}
\end{equation}
which is exactly the same as the case of the free theory, \eqref{scalarconstr}.
By varying the action with respect to the scale factor, we find
\begin{equation}
   4 \dot{H} \left(\xi  \,\Theta+\Mpl^{2}\right)= 
-4 \ddot{\Theta} \xi -4 H \dot{\Theta} \xi +\left(-6 \xi  \,H^{2}+m^{2}\right) \Theta-6 \Mpl^{2} H^{2}+2 \Lambda \,\Mpl^{2}-\dot{\chi}^{2}+2 U \! \left(\chi \right)\,.
\label{ddawithmatter}
\end{equation}
By setting $\sigma=0$, and substituting the constraint equation for $H^2$, we recover the standard acceleration equation,
\begin{equation}
    \dot{H}=-\frac{1}{2\Mpl}(\varepsilon_{\chi}+p_{\chi}),\quad p_{\chi}=\frac{1}{2}\dot{\chi}^2-U(\chi)\,.
\end{equation}
Similarly, even in the case $\sigma^2=\Theta\neq0$, \eqref{ddawithmatter}
is automatically satisfied if \eqref{constraintMatter}, \eqref{sigmaconstr} and the scalar field equation,
\begin{equation}
    \ddot{\chi}+3H\dot{\chi}+V_{,\chi}=0\,,
\end{equation}
are both satisfied. Thus the system can be consistently solved by solving the three equations, the Hamiltonian and the scalar constraints, and the field equation for $\chi$.

We can see that since the matter is minimally coupled to the scalar, it cannot affect the evolution of space-time, characterized by the Hubble parameter. Rather, the evolution of the Hubble parameter is determined independently, only in terms of the mass and coupling associated with the constrained scalar. 
This means that in the case of a perfect fluid, we are to solve the following equations: 
\begin{equation}
    \begin{split}
    \dot{H}(t)&=\frac{12 \xi  \,H(t)^{2}-m^{2}}{-6 \xi}\,,\\
        \dot{\Theta}(t) &= 
\frac{-6 H^{2} \Theta \xi -6 \Mpl^{2}H^{2}+m^{2} \Theta+2 \varepsilon}{6H \xi}\,,\\
\dot{\varepsilon}(t)&=-3H(t)(\varepsilon(t)+p(t))\,.
    \end{split}
\end{equation}

The Hubble parameter is fully determined by the first equation. It is easy to see how this reflects on the evolution of the perfect fluid. 

For the equation of state, $p=w\varepsilon$ with a constant $w$,
the energy density is proportional to $a^{-3(1+w)}$,
\begin{equation}
    \varepsilon\propto a^{-3(1+w)}\,. 
\end{equation}
The Hubble parameter and the scale factor satisfy the same equation as in the case when the matter was absent, which is given in the subsection \ref{section::HubP}. 
As a special case, let us consider the case when the Hubble parameter is constant,
\begin{equation}
    H=\frac{m}{2\sqrt{3\xi}}\,.
\end{equation}
This leads to the energy density given by
\begin{equation}
    \varepsilon=\varepsilon_0e^{-3H(1+w)t}\,,
\end{equation}
where $\varepsilon_0$ is the energy density at $t=0$, and the constrained scalar by

\begin{align}
 & \Theta =c_{\Theta}  {\mathrm e}^{\frac{m t}{2\sqrt{3\xi}}}+\frac{\Mpl^2}{\xi}- \frac{4\varepsilon_0}{ m^2(4+3w)}e^{-\frac{\sqrt{3}m}{2\sqrt{\xi}}(1+w)t}
 \\
 &\quad = c_\Theta a(t)+\frac{\Mpl^2}{\xi}-\frac{4}{m^2(4+3w)}\varepsilon(t)\,,
\end{align}
where $ c_{\Theta}$ is a constant of integration. 

Therefore, we can see that see that the external matter contributes to the evolution of the scalar field, but it nevertheless does not affect the cosmological background.

\subsection{The non-minimal coupling with the constrained scalar}
\label{sec:nmcouplematter}
In the previous subsection, we have seen that if matter is minimally coupled to both scalar and gravity, it does not affect the evolution of space-time if the constrained scalar is non-vanishing. (Otherwise, if $\sigma=0$, everything evolves according to the standard cosmology.) This behaviour is curious, but in direct tension with General Relativity due to nucleosynthesis. In this subsection, we will resolve this issues, and show that the external matter can nevertheless influence cosmological background through the non-minimal coupling with the constrained scalar field. In particular, we can consider the following action: 
\begin{equation}\label{nminconstrscact}
    S=\int d^4x\sqrt{-g}\left(\frac{\Mpl^2}{2}(R-2\Lambda)+\sigma^2\left(-\frac{m^2}{2}+\frac{\xi}{2}R\right)+f(\sigma)L_m\right)\,.
\end{equation}
where $L_m$ stands for the matter Lagrangian. Here, $f(\sigma)$ is a general function of the constrained scalar. In this work, we will consider its simplest form, which is given by: 
\begin{equation}\label{simplestnmincou}
    f\left(\sigma\right)=\lambda\frac{\sigma^2}{\Mpl^2}\,,
\end{equation}
where $\lambda$ is the dimensionless coupling.
Hence, 
\begin{equation}
    S=\int d^4x\sqrt{-g}\left(\frac{\Mpl^2}{2}(R-2\Lambda)+\sigma^2\left(-\frac{m^2}{2}+\frac{\xi}{2}R+ \bar{L}_m\right)\right)\,,
\end{equation}
where $\bar{L}_m=\lambda L_m/\Mpl^2$, which has mass dimension two, for convenience.  
As for the matter Lagrangian, $L_m$, for a scalar field, we have 
\begin{equation}
    {L}_m=-\frac{1}{2}g^{\mu\nu}\partial_{\mu}\chi\partial_{\nu}\chi-U(\chi)\,,
\end{equation}
for a vector field,
\begin{equation}
     {L}_m=-\frac{1}{4}F_{\mu\nu}F^{\mu\nu}\,,
\end{equation}
and for a perfect fluid,
\begin{equation}
{L}_m=p\,,
\end{equation}
where $p$ is the pressure. Below we consider the cases of a scalar and a perfect fluid.

First, let us consider the case of a scalar field. The energy density and the pressure on the FLRW background is given by
\begin{equation}
    \varepsilon=\frac{1}{2}\dot{\chi}^2+U\,,\quad  p=\frac{1}{2}\dot{\chi}^2-U\,.
\end{equation}
Then, the Hamiltonian constraint gives 
\begin{equation}\label{ommmatter}
    \dot{\Theta}=\frac{-6 H^{2} \Theta \xi -6 \Mpl^{2} H^{2}+2\Theta \bar{\varepsilon}+m^{2} \Theta+2 \Lambda \,\Mpl^{2}}{6   H \xi}\,,
\end{equation}
where $\bar{\varepsilon}=\lambda\varepsilon/\Mpl^2$ and $\Theta=\sigma^2$.
By varying the action with respect to $\sigma$, we find
\begin{equation}
    \sigma  \left(-12 \xi  \,H^{2}-6 \dot{H} \xi +m^{2}+\frac{2 V(\chi)-\dot{\chi}^2}{\Mpl^2}\right)=0\,.
\end{equation}
Assuming $\sigma\neq0$ again, this gives the equation for the Hubble parameter,
\begin{equation}\label{hubbmatter}
    \dot{H} = 
-\frac{12 \xi  \,H^{2}-m^{2}+2\bar{p}}{6 \xi}\,,
\end{equation}
where $\bar{p}=\lambda p/\Mpl^2$. 
Thus, we can see that if we couple the matter non-minimally to the scalar field (but minimally to gravity), it can now contribute to the evolution of space-time. While it seems that the above equation is independent of $\sigma$ as it was in the minimal case, we have to be careful. 
This is because the equation of motion for the scalar (and thus the continuity equation upon the above identification) becomes modified, 
\begin{equation}
 \ddot\chi+\left(3 H + \frac{\dot{\Theta}}{\Theta}\right) \dot{\chi}+V_{,\chi}=0\,.
\label{chimattereq}
\end{equation}
In the language of the energy conservation, this modification implies 
\begin{equation}
  \dot{\varepsilon}+\left(3H+\frac{\dot{\Theta}}{\Theta}\right)(\varepsilon+p)=0\,.
  \label{Emattereq}
\end{equation}
Finally, by varying with respect to the scale factor, we find the acceleration equation, 

\begin{equation}
    \Theta\left(\frac{2U(\chi)-\dot{\chi}^2}{\Mpl^2}-6 \xi  \,H^{2}+m^{2}-4 \dot{H} \xi \right)-2\xi\left(\ddot\Theta+2H\dot{\Theta}\right)-6 \Mpl^{2} H^{2} +2 \Mpl^{2} \left(\Lambda -2 \dot{H}\right)=0\,,
\end{equation}
which is automatically satisfied, provided that \eqref{ommmatter}, \eqref{hubbmatter} and \eqref{chimattereq} hold.

Let us now consider a fluid with the equation of state $p=w \varepsilon$. 
Then, \eqref{Emattereq} can be integrated to give
\begin{equation}
    \varepsilon=\varepsilon_0\left[\left(\frac{a_0}{a}\right)^{3}\left(\frac{\Theta_0}{\Theta}\right)\right]^{(1+w)},
    \label{energywithm}
\end{equation}
where $a_0$ and $\Theta_0$ are the initial values of the scale factor and the constrained scalar, and $\varepsilon_0$ is the initial energy density. 
This expression can be substituted into the equations for the Hubble parameter and the constrained scalar to obtain
\begin{align}
    %\begin{split}
          \dot{H} =&
-\frac{12 \xi  \,H^{2}-m^{2}+2w\, \bar{\varepsilon}}{6 \xi}\,,
\label{dotHwithm}\\
  \dot{\Theta}=&\frac{(m^2+2\bar{\varepsilon}-6\xi H^{2} )\Theta -6 \Mpl^{2} H^{2}+2 \Lambda \,\Mpl^{2}}{6   H \xi}\,.
\label{dotThwithm}    %\end{split}
\end{align}
In the following, we will solve these equations for several different values of $w$. 
For simplicity, we set 
\begin{equation}
    m=1\,,\quad \Lambda=1/3\,,\quad \varepsilon_0=10\,,\quad \xi=1\,, \quad\lambda=1
\end{equation}
in the reduced Planck units $\Mpl=1$.
We will compare thus obtained solutions with those in the free case. In the figures shown below, {the blue thick lines correspond to the cases with matter coupling and the pink dotted lines to those without matter, i.e., the free case. }

\begin{center}
    \textit{Case 1: p=0 (dust)}
\end{center}
First, let us consider the case of dust, that is, $w=0$. In this case, we immediately notice that the equation for the Hubble parameter becomes independent of the energy density,
\begin{equation}
    \dot{H} =-\frac{12 \xi  \,H^{2}-m^{2}}{6 \xi}\,.
\end{equation}
We have already explored the solutions to this equation, with the constrained scalar evolution being modified. 
In particular, while the Hubble parameter and the scale factor are identical to those in the free case without matter, the time derivative of the constrained scalar acquires an additional boost from the energy density $\propto \varepsilon\Theta$, as seen in \eqref{dotThwithm},
which makes it evolve to large values faster than the free case. 
This behavior is presented in Fig.~\ref{fig:dust}.
\begin{figure}[h!]
    \centering
    \includegraphics[width=7cm]{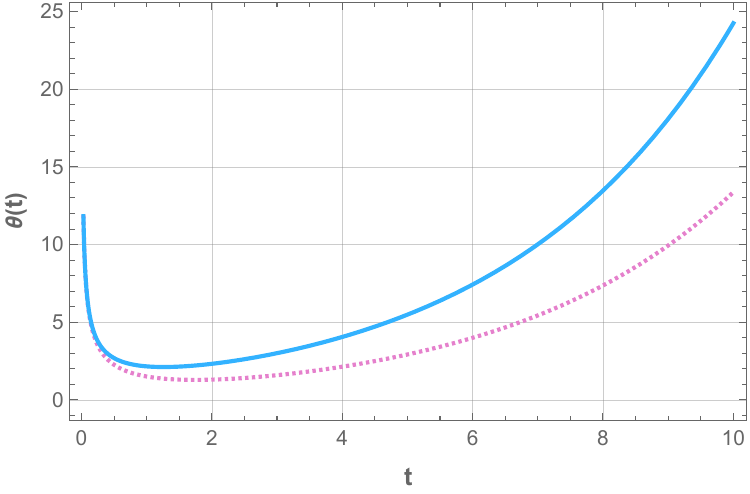}
    \caption{ \textbf{(Jordan frame)} Evolution of the constrained scalar coupled with a dust fluid $w=0$. Initial conditions are given by  $H(t_i)= 10$, $\Theta_0= \Theta(t_i)=9$, where $t_i\approx 0.05$ using the asymptotic behavior
$H\approx1/(2t)$ for $w<1$. The blue line corresponds to the case with matter, while the pink dotted one corresponds to the free case.  
    } 
\end{figure}
\label{fig:dust}

\begin{figure}[h!]
    \centering

    \begin{minipage}{0.3\textwidth}
        \centering
        \includegraphics[width=\linewidth]{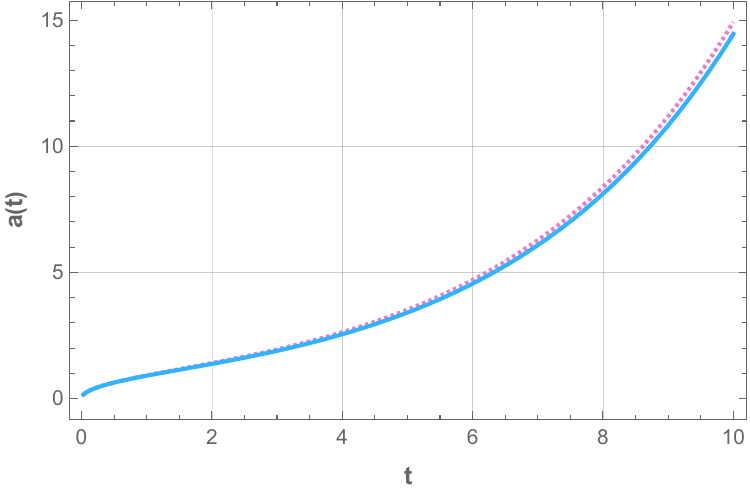 }
    \end{minipage}
    \hfill
    \begin{minipage}{0.3\textwidth}
        \centering
        \includegraphics[width=\linewidth]{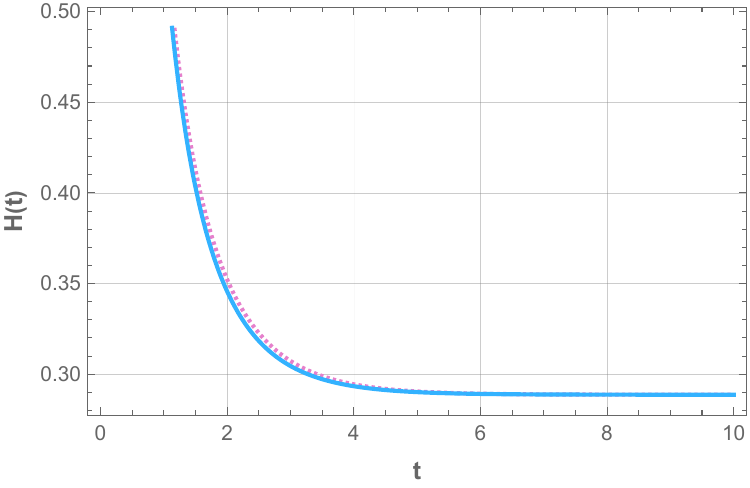 }
    \end{minipage}
    \hfill
    \begin{minipage}{0.3\textwidth}
        \centering
        \includegraphics[width=\linewidth]{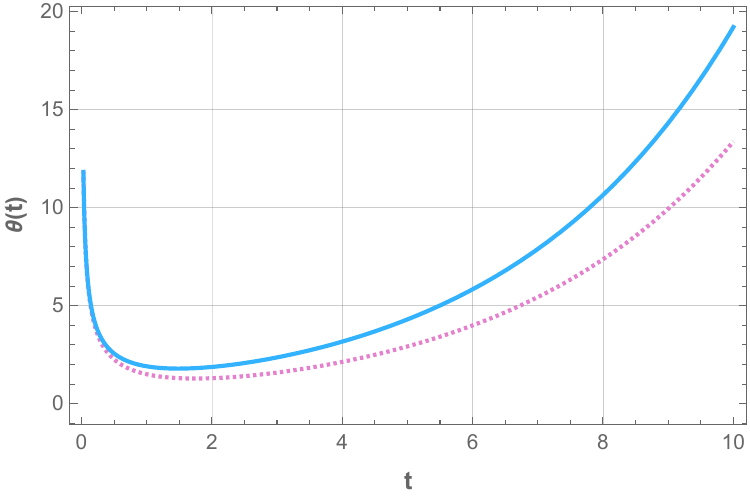 }
    \end{minipage}

    \caption{\textbf{(Jordan frame)} Solution to the equations (\ref{dotHwithm}) and (\ref{dotThwithm}) with the radiation equation of state $w=1/3$ for the constrained scalar. Initial conditions are given by  $H(t_i)= 10$, $\Theta_0= \Theta(t_i)=9$, where $t_i\approx 0.05$ using the asymptotic behavior
$H\approx1/(2t)$ for $w<1$. 
    %Initial conditions are given by  $a(0.2)=0.2, H(0.2)=10, \Theta(0.2)=9$. 
} 
    \label{fig:radiation}
\end{figure}

\begin{center}
    \textit{Case 2: $p=\dfrac{1}{3}\varepsilon$ (radiation)}
\end{center}
Next, let us consider the $w=1/3$ case. The resulting behaviors of the scale factor $a(t)$, the Hubble parameter $H(t)$, and the constrained scalar $\Theta(t)$ are shown in Fig.~\ref{fig:radiation}.
Except for $\Theta$, we find that the solution behaves very similar to Case 2 of the free case with $\alpha=0$. 
This may be explained as follows. 
From \eqref{energywithm}, we have $\varepsilon\propto a^{-4}\Theta^{-4/3}$. 
From \eqref{dotThwithm}, one expects $\Theta\propto a^{-1}$ in the limit $a\to0$.
Thus $\varepsilon\propto a^{-3}$. 
With this behavior of $\varepsilon$, \eqref{dotHwithm} implies $H\propto a^{-2}$ as $a\to0$. 
Plugging these back into \eqref{dotHwithm} and \eqref{dotThwithm}, one finds these behaviors are consistent. 
Namely, as long as one assumes a diverging $H$ in the limit $t\to0$, the presence of the matter energy density $\varepsilon$ plays a negligible role in this limit. 
Then as the solution eventually approaches the accelerating universe with a constant $H=m/(2\sqrt{3\xi})$, where $\varepsilon$ is again negligible, the solution approaches the one without matter in both limits $t\to0$ and $t\to\infty$.

\begin{figure}[h!]
    \centering

    \begin{minipage}{0.3\textwidth}
        \centering
        \includegraphics[width=\linewidth]{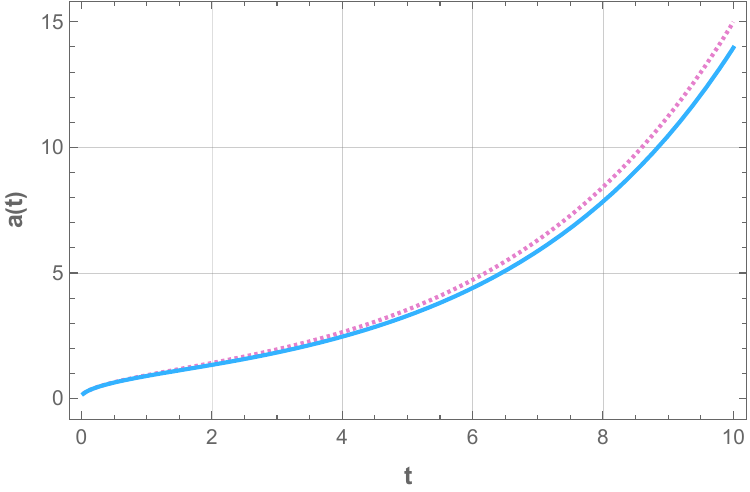}
    \end{minipage}
    \hfill
    \begin{minipage}{0.3\textwidth}
        \centering
        \includegraphics[width=\linewidth]{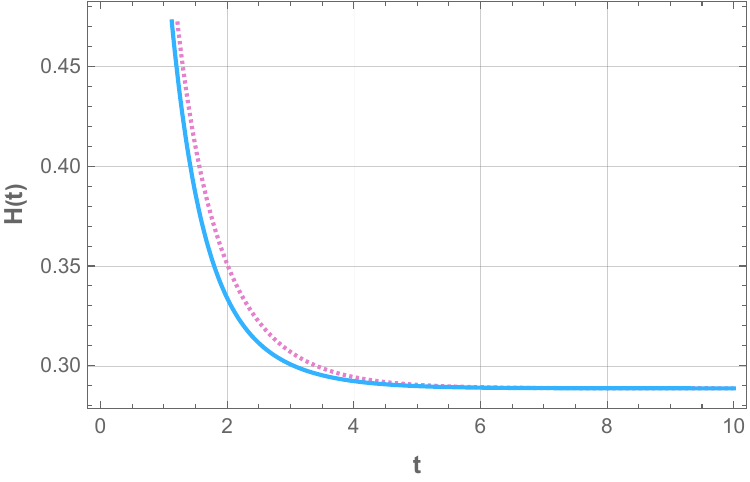}
    \end{minipage}
    \hfill
    \begin{minipage}{0.3\textwidth}
        \centering
        \includegraphics[width=\linewidth]{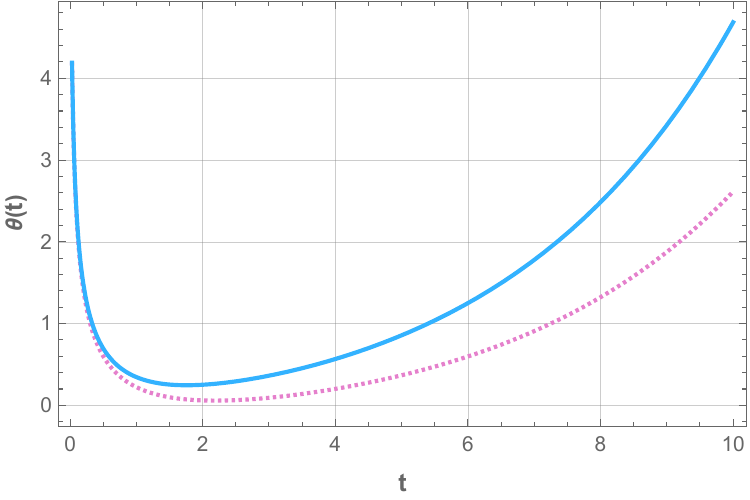}
    \end{minipage}
     \begin{minipage}{0.3\textwidth}
        \centering
        \includegraphics[width=\linewidth]{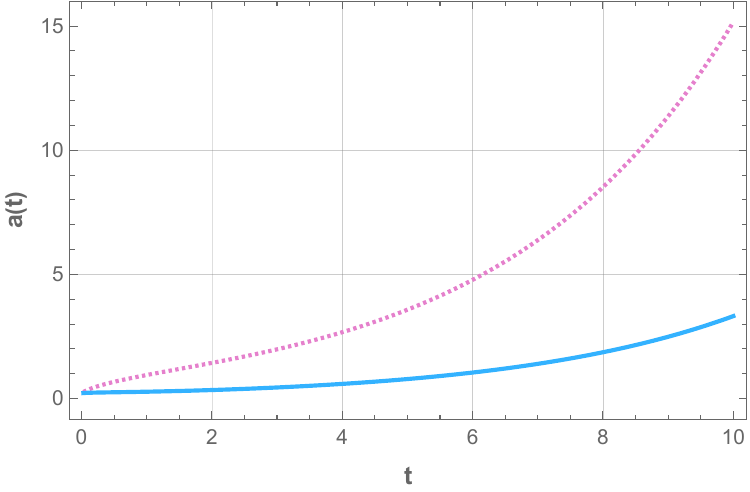}
    \end{minipage}
    \hfill
    \begin{minipage}{0.3\textwidth}
        \centering
        \includegraphics[width=\linewidth]{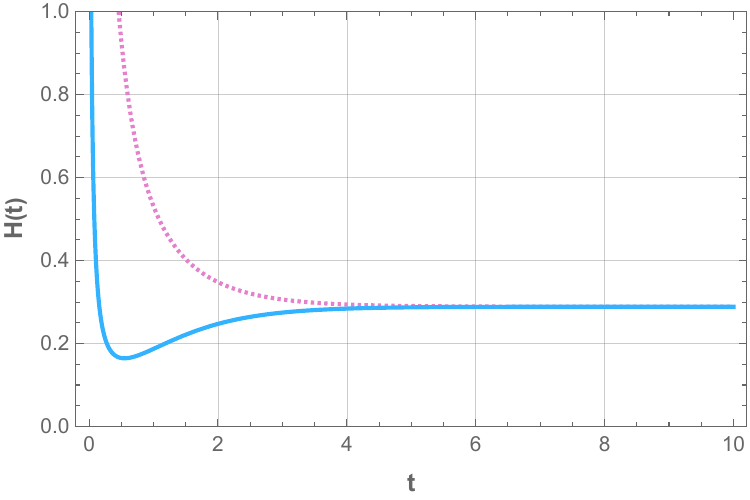}
    \end{minipage}
    \hfill
    \begin{minipage}{0.3\textwidth}
        \centering
        \includegraphics[width=\linewidth]{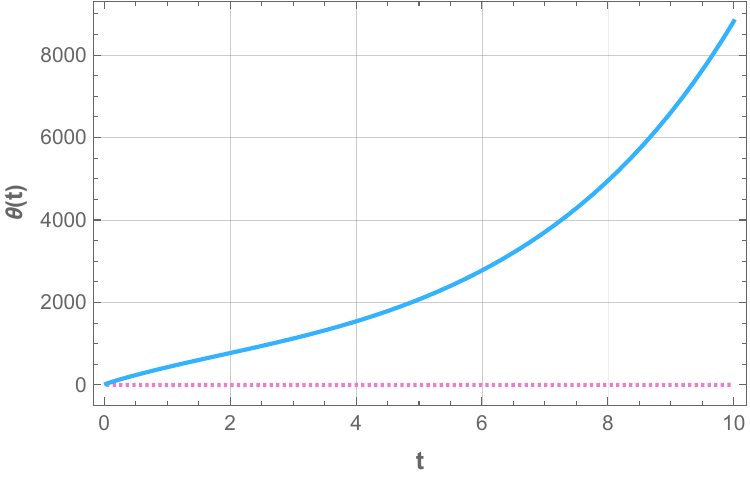}
    \end{minipage}

    \caption{ \textbf{(Jordan frame)} Solution to the equations (\ref{dotHwithm}) and (\ref{dotThwithm}) with ultra-hard equation of state $\omega=1$. 
    %Initial conditions are given by  $a(0.2)=0.2, H(0.2)=10, \Theta(0.2)=4$. 
    Initial conditions are determined by $H(t_i)=10$ and $\Theta_0=\Theta(t_i)=4$.
    The asymptotic behavior at $t_i\to0$ is given by  $H\approx1/((2-c)t)$ if $c\ll1$ and $H\approx1/(ct)$ if $c\gg1$, where $c=\varepsilon_0/(3\xi H(t_i)^2)$. 
    The figures in the upper panel correspond to the case with $\varepsilon_0=10$ ($c=1/30$). 
    Those in the lower panel correspond to the case $\varepsilon_0=6000$ ($c=200$). } 
    \label{fig:w=1}
\end{figure}

\begin{center}
    \textit{Case 3: $p=\varepsilon$ (ultra-hard equation of state)}
\end{center}
Let us now consider the case $\omega=1$, with the ultra-hard equation of state. In this case, we find the evolutionary behavior as presented in Fig.~\ref{fig:w=1}.
As one can see, in the upper example, where the $\varepsilon$
term in the equations for (\ref{dotThwithm}) is small in comparison with the
$H^2$ term, there is again only a small difference in the evolution of
$H$ from the free case, except that a small suppression due to the
$\varepsilon$ term relative to the free case seems a bit more apparent
when the evolution transitions from decelerated expansion into
accelerated expansion, which leads to a larger delay in time in the
growth of the scale factor. In the lower example, where the
$\varepsilon$ term dominates in the beginning, one can show that
$H\times\Theta\approx const.$, and $H$ decreases rapidly until the
$\varepsilon$ term becomes comparable to the $m^2$ term. After that,
$H$ experiences the super-Hubble expansion because $\dot H\sim
m^2/\xi>0$. The behavior shown in the figure is consistent with this
qualitative estimate.

\begin{center}
    \textit{Case 4: $p=-\dfrac{2}{3}\varepsilon$ }
\end{center}

\begin{figure}[h]
    \centering

    \begin{minipage}{0.3\textwidth}
        \centering
        \includegraphics[width=\linewidth]{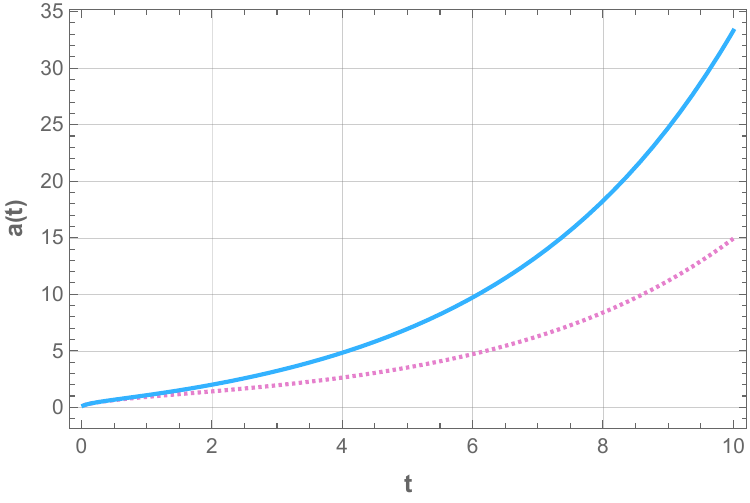}
    \end{minipage}
    \hfill
    \begin{minipage}{0.3\textwidth}
        \centering
        \includegraphics[width=\linewidth]{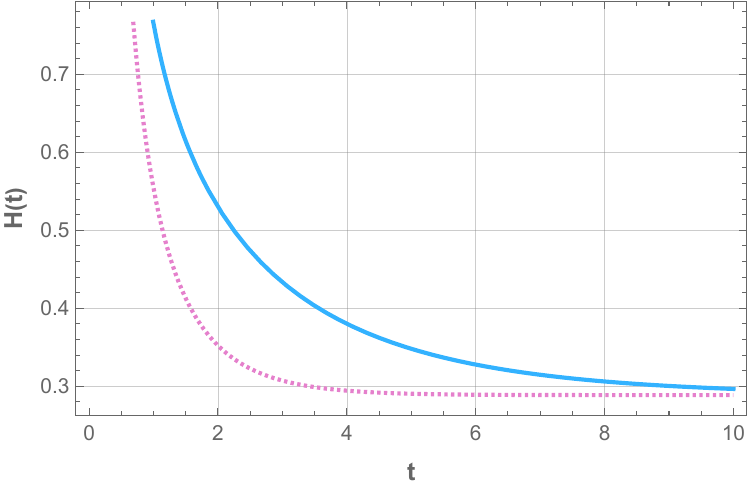}
    \end{minipage}
    \hfill
    \begin{minipage}{0.3\textwidth}
        \centering
        \includegraphics[width=\linewidth]{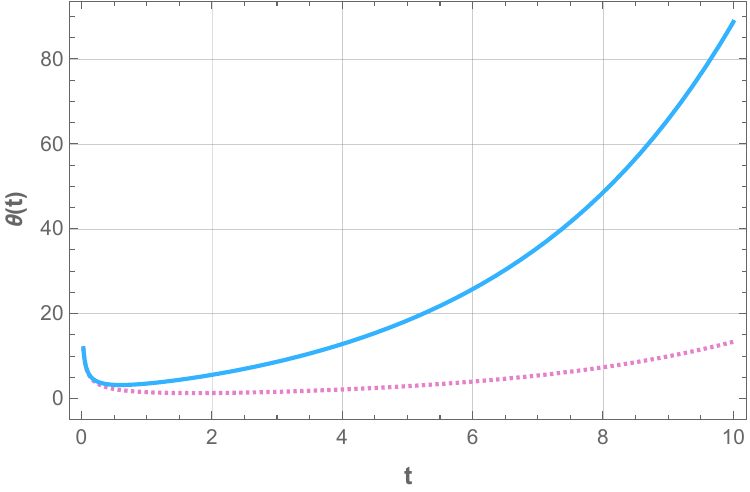}
    \end{minipage}

    \caption{ \textbf{(Jordan frame)} Solution to the equations (\ref{hubbmatter}) and (\ref{ommmatter}) with  equation of state $\omega=-2/3$. Initial conditions are given by  $H(t_i)= 10$, $\Theta_0= \Theta(t_i)=9$, where $t_i\approx 0.05$ using the asymptotic behavior $H\approx1/(2t)$ for $w<1$.    
} 
    \label{fig:w=-2/3}
\end{figure}

We have seen in the above that the suppression in the Hubble parameter around the transition time is larger for larger $w$, and that there is no suppression for $w=0$. This suggests that there may be an enhancement in the Hubble parameter for $w<0$, and it is larger for a larger absolute value of $w$. To see this, we consider $w=-2/3$. 
The result is presented in Fig.~\ref{fig:w=-2/3}.
We can indeed notice that the Hubble parameter is enhanced in comparison with the free case.

\begin{center}
    \textit{  Case 5: $p=-¥\varepsilon$ (cosmological constant)}
\end{center}

\begin{figure}[h]
    \centering

    \begin{minipage}{0.3\textwidth}
        \centering
        \includegraphics[width=\linewidth]{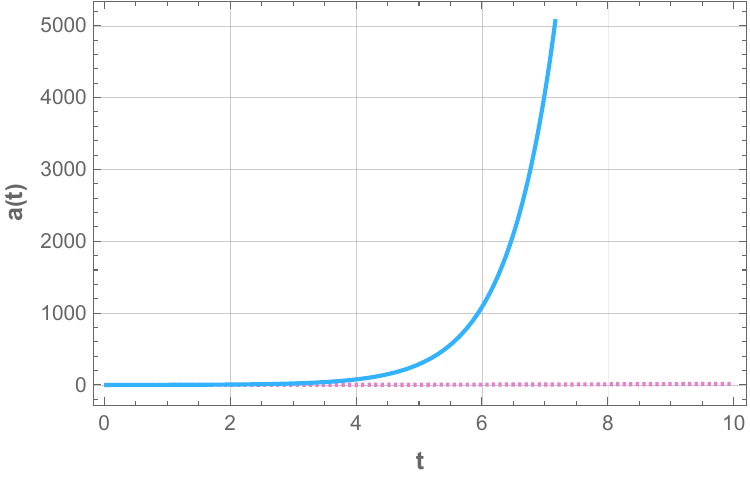}
    \end{minipage}
    \hfill
    \begin{minipage}{0.3\textwidth}
        \centering
        \includegraphics[width=\linewidth]{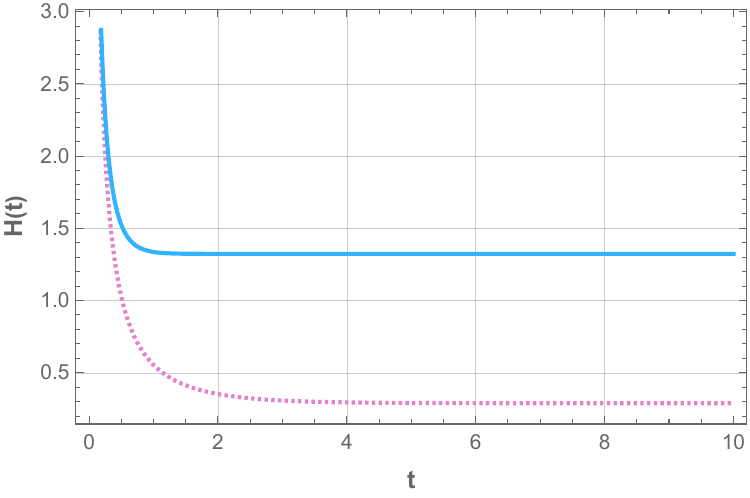}
    \end{minipage}
    \hfill
    \begin{minipage}{0.3\textwidth}
        \centering
        \includegraphics[width=\linewidth]{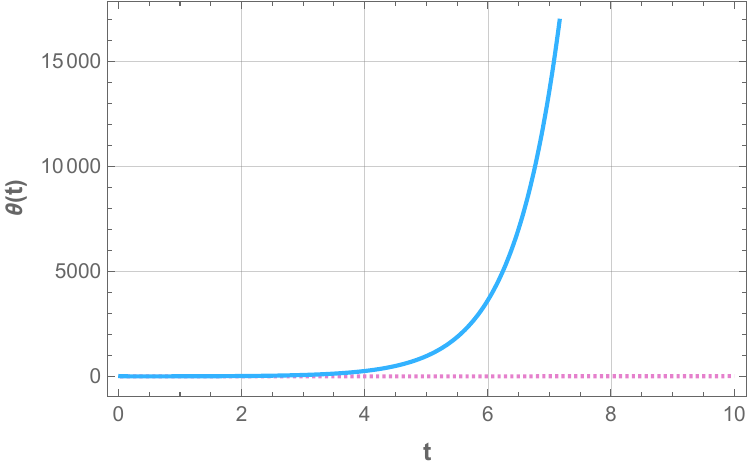}
    \end{minipage}

    \caption{ \textbf{(Jordan frame)} Solution to the equations (\ref{hubbmatter}) and (\ref{ommmatter}) with  equation of state $\omega=-1$. Initial conditions are given by  $H(t_i)= 10$, $\Theta_0= \Theta(t_i)=9$, where $t_i\approx 0.05$ using the asymptotic behavior $H\approx1/(2t)$ for $w<1$.   
} 
    \label{fig:w=-1}
\end{figure}

Finally we consider the case $w=-1$. In this case, as clear from \eqref{dotHwithm}, the presence of matter is equivalent to replacing $m^2$ by $m^2+2\bar{\varepsilon}_0$. 
This implies that $H$ approaches a constant larger than the one in the free case in the limit $t\to\infty$. 
As can be seen from \eqref{dotThwithm}, the same effect applies also to $\Theta$. 
The resulting behaviors of $a$, $H$, and $\Theta$ are shown in Fig.~\ref{fig:w=-1}. 
As one can see, the above expectation is indeed confirmed. 

Overall, we confirm the following trend. In the case of dust, $w=0$ the the solution matches with the free case.
In the non-minimal case, the Hubble parameter is suppressed around the time of transition from the decelerating phase to accelerating phase for $w>0$,
while it is enhanced for negative $w$. But the Hubble parameter approaches that in the free case in both limits $t\to0$ and $t\to\infty$, for $-1<w\leq1$.
The $w=-1$ case is special in that it is equivalent to replacing $m^2$ by $m^2+\bar{\varepsilon}_0$.

\begin{center}
    \textit{Case 6: Super-Hubble expansion}
\end{center}

\begin{figure}[h]
    \centering

    \begin{minipage}{0.3\textwidth}
        \centering
        \includegraphics[width=\linewidth]{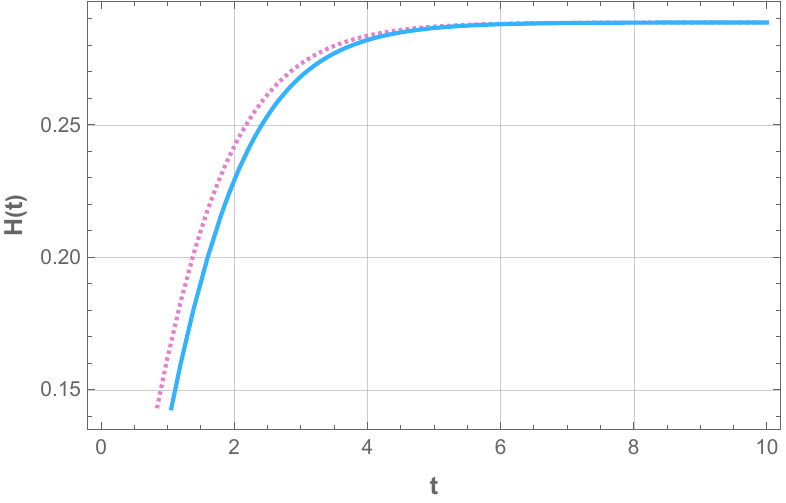}
    \end{minipage}
    \hfill
    \begin{minipage}{0.3\textwidth}
        \centering
        \includegraphics[width=\linewidth]{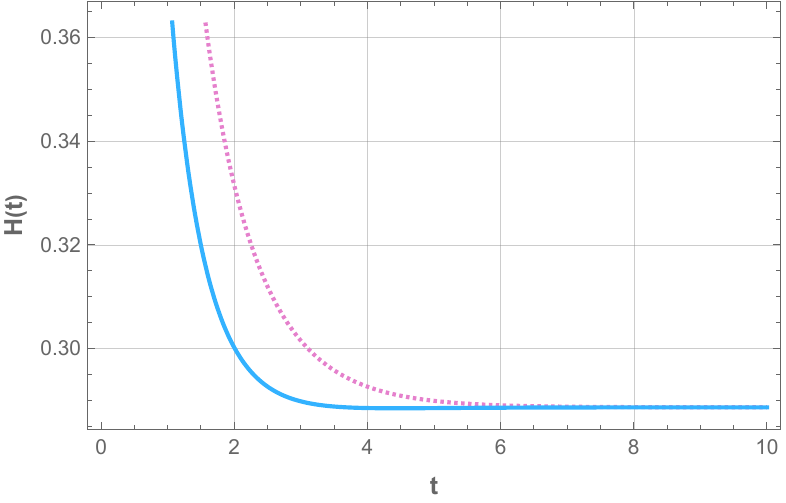}
    \end{minipage}
    \hfill
    \begin{minipage}{0.3\textwidth}
        \centering
        \includegraphics[width=\linewidth]{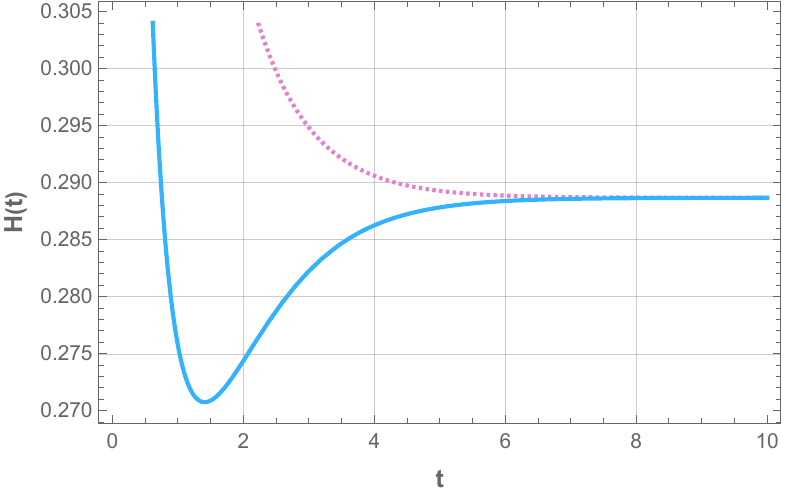}
    \end{minipage}
 \centering
\begin{minipage}{0.3\textwidth}
        \centering
        \includegraphics[width=\linewidth]{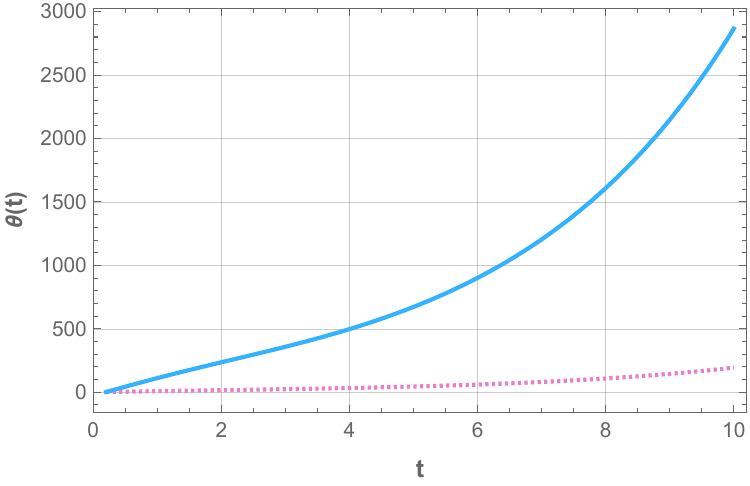}
    \end{minipage}
    \hfill
    \begin{minipage}{0.3\textwidth}
        \centering
        \includegraphics[width=\linewidth]{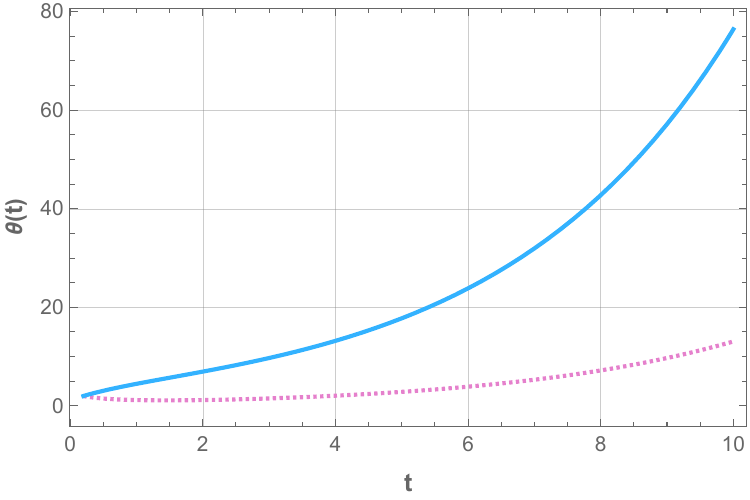}
    \end{minipage}
    \hfill
    \begin{minipage}{0.3\textwidth}
        \centering
        \includegraphics[width=\linewidth]{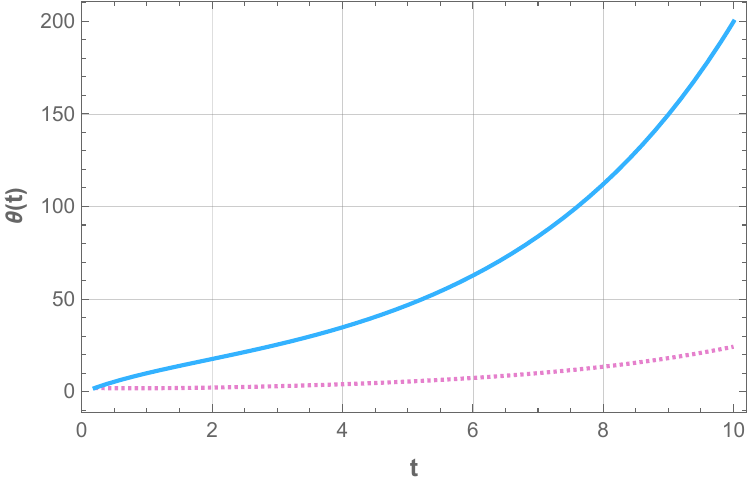}
    \end{minipage}

    \caption{\textbf{(Jordan frame)} The dependence on the initial Hubble parameter. The larger $H$ corresponds to a larger amount of matter at the initial time.
    All the figures are for the same initial value of the scalar field, $\Theta(t_i)=2$, while for the initial Hubble parameter, the left figures are with $H(t_i)=0.005$ which exhibits only the supper-Hubble behavior, the middle ones with $H(t_i)=1$ which exhibits initially decelerating and eventually accelerating behavior, and the right ones with $H(t_i)=0.5$ which exhibits super-Hubble behavior between the initial decelerated and final accelerated stages. }
    \label{fig:superH}
\end{figure}

We should mention that the above behavior holds for different choices of parameters as well as for a wide range of initial conditions. 
However, with different choices of the initial conditions, it is also possible to arrive at the supper-Hubble expansion, that relaxes to an exponential expansion, which we have encountered in Case 1 in the free case, with $\beta=0$. 
For example, in Fig.~\ref{fig:superH}, we present the results for the case $w=1/3$ with the same parameters but for three different initial conditions.  

\subsection{The effective equation of state}

In the previous subsections, we have analyzed the evolution of the Hubble parameter and the constrained scalar for various choices of external matter. However, one might want to know what kind of equation of state these dynamics correspond to. 
 The usual approach would be to consider the acceleration equation and the constraint equation, identify the effective energy density and pressure by identifying the modified Friedman equations with the standard ones, and define the effective equation of state as the ratio of the two. However, in our case, this would imply that the constrained scalar enters the equations of state even if the theory is free. This does not make too much sense, as it does not affect the evolution of spacetime. 
 Therefore, we define the effective equation of state by identifying the formula given by the acceleration of the expansion in terms of the equation of state which holds in the standard Einstein theory with a fluid.
 Explicitly, we define the effective equation of state by 
\begin{equation}\label{omegaS}
    w_S=-1-\frac{2\dot{H}}{3H^2}\,,
\end{equation}
In this subsection, we will substitute the previously found expressions for the Hubble parameter, and evaluate $w_S$. 

\subsubsection{The free case}

Let us first consider the case without matter. Then, the Hubble parameter is given by
\begin{equation}
    \dot{H}=\frac{1}{6\xi}\left[-12\xi H^2+m^2\right]\,.
\end{equation}
By substituting this into (\ref{omegaS}), we find
\begin{equation}
    w_S=\frac{1}{3}-\frac{m^2}{9\xi H^2}\,.
\end{equation}
Therefore, for large values of the Hubble parameter, $ w_S$ behaves as an equation of state for radiation. 
This corresponds to Case 2 discussed previously.
As the Hubble parameter decreases, $w_S$ crosses zero, and eventually approaches $-1$ as the Hubble parameter approaches a constant given by $H=m/(2\sqrt{3\xi})$.

\subsubsection{In the presence of matter}

\begin{figure}[h]
    \centering

    \begin{minipage}{0.45\textwidth}
        \centering
        \includegraphics[width=\linewidth]{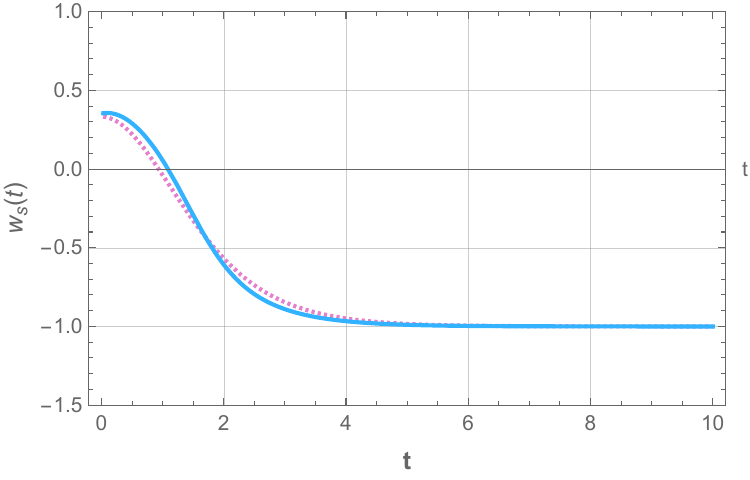}
    \end{minipage}
    \hfill
    \begin{minipage}{0.45\textwidth}
        \centering
        \includegraphics[width=\linewidth]{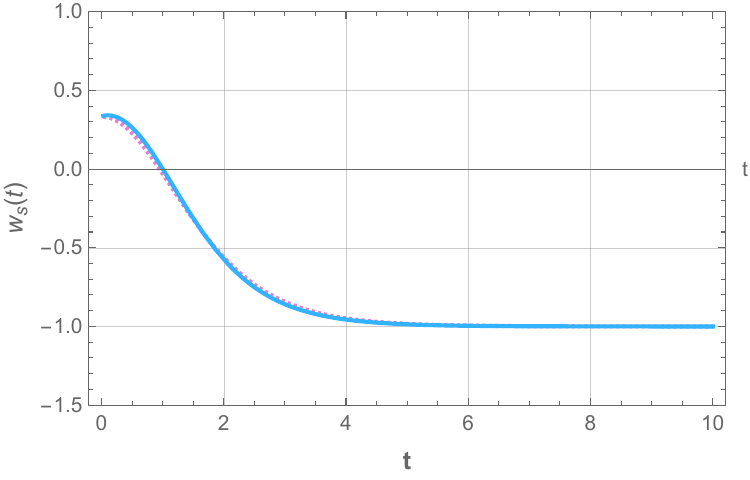}
    \end{minipage}
    \centering
    \begin{minipage}{0.45\textwidth}
        \centering
        \includegraphics[width=\linewidth]{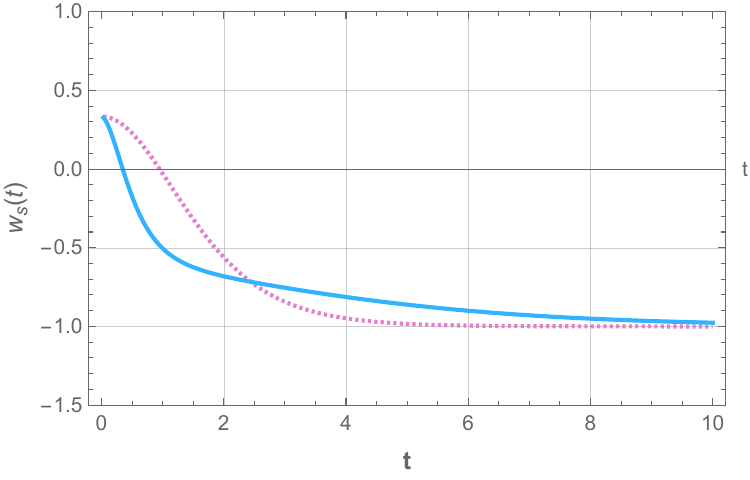}
    \end{minipage}
    \hfill
 \begin{minipage}{0.45\textwidth}
        \centering
        \includegraphics[width=\linewidth]{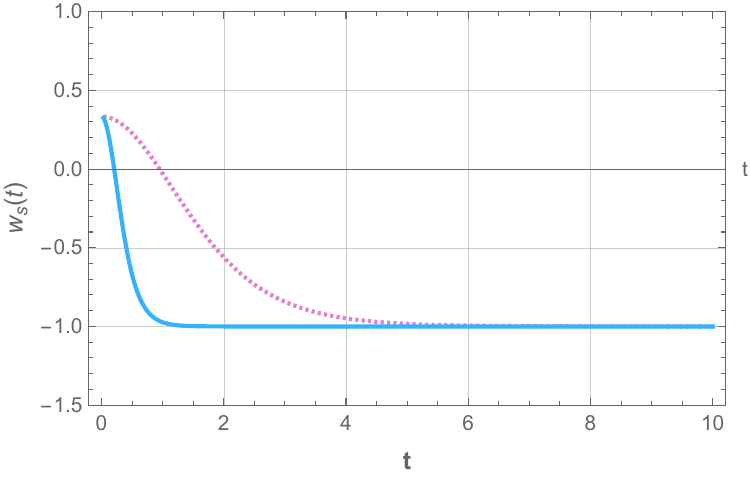}
    \end{minipage}
    \caption{ \textbf{(Jordan frame)} The evolution of the effective equation of state with matter. 
    The upper left graph is for $w=1$, the upper right for $w=1/3$, the lower left for $w=-2/3$, and the lower right for $w=-1$. 
    The parameters and the initial conditions are the same as those discussed in Subsection \ref{sec:nmcouplematter}, corresponding to Cases 2 to 5. For comparison, we have also included the pink dashed lines, which correspond to the cases without matter (free case).  }
    \label{fig:wsCases}
\end{figure}

\begin{figure}[h]
    \centering

    \begin{minipage}{0.3\textwidth}
        \centering
        \includegraphics[width=\linewidth]{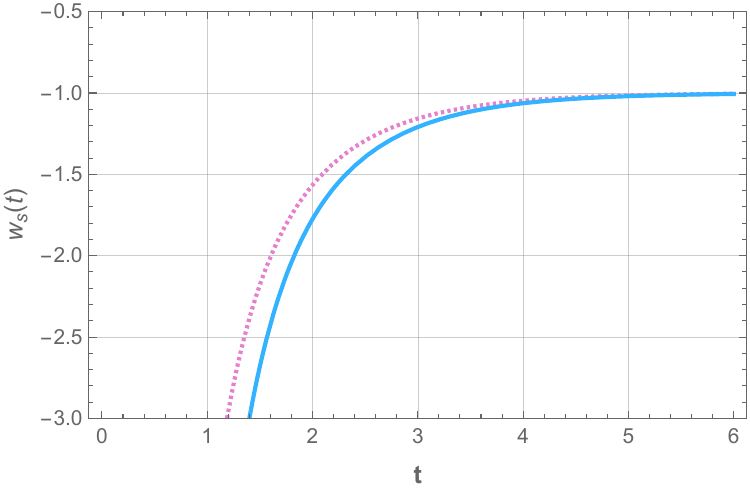}
    \end{minipage}
    \hfill
    \begin{minipage}{0.3\textwidth}
        \centering
        \includegraphics[width=\linewidth]{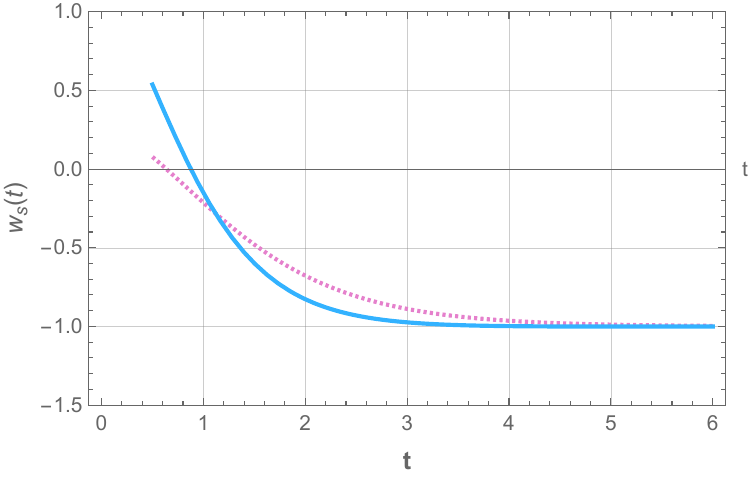}
    \end{minipage}
    \hfill
    \begin{minipage}{0.3\textwidth}
        \centering
        \includegraphics[width=\linewidth]{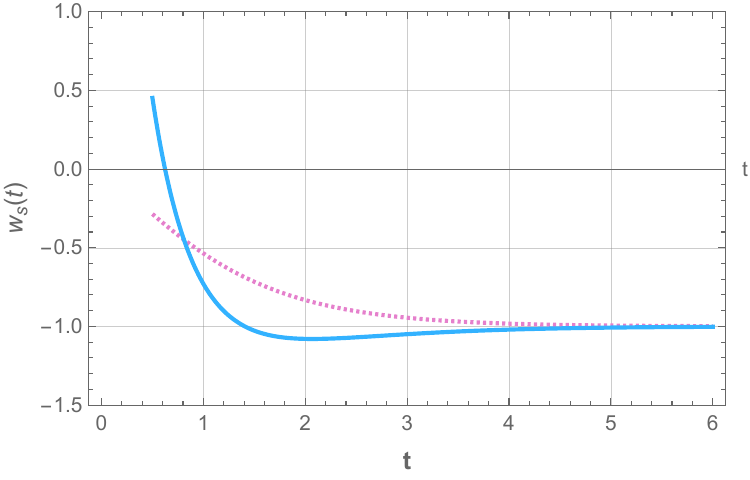}
    \end{minipage}

    \caption{ \textbf{(Jordan frame)} The evolution of the effective equation of state with radiation, $w=1/3$. 
    The parameters are the same as Fig.~\ref{fig:wsCases}, but with different initial conditions. 
    The left graph is for the initial conditions $H(t_i)=0.05$ and $\Theta(t_i)=2$, the middle one is for $H(t_i)=1$ and $\Theta(t_i)=2$, and the right one is for $H(t_i)=0.5$ and $\Theta(t_i)=2$.}
    \label{phantomnm1}
\end{figure}

Let us now consider the effect of matter. The effective equation of state is modified to
\begin{equation}
    w_S=\frac{1}{3}-\frac{m^2}{9\xi H^2}+\frac{2w\,\bar{\varepsilon}}{9\xi H^2}\,.
\end{equation}
In the following, we will plot the values of the equation of state and compare it with the free case. In Figs.~\ref{fig:wsCases} we present the results corresponding to the previously considered cases of $w=1$, $w=1/3$,  $w=-2/3$, and $w=-1$, with initial conditions given in Cases 2, 3, 4, and 5. In Fig.~\ref{phantomnm1}, we present the result corresponding to Case 6, i.e., the case in which the universe experience a super-Hubble expansion phase. 
Interestingly, for the latter cases that encompass the super-Hubble expansion we find that values smaller than $-1$ as well, given in the Figure \ref{phantomnm1}.

\begin{figure}[h]
    \centering

    \begin{minipage}{0.3\textwidth}
        \centering
        \includegraphics[width=\linewidth]{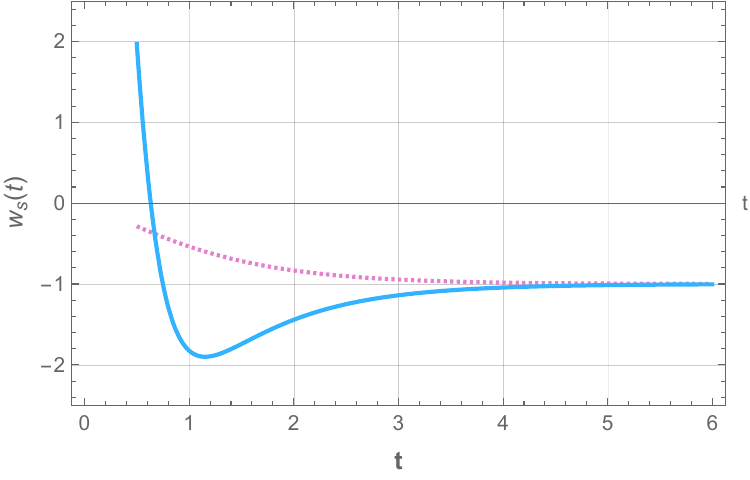}
    \end{minipage}
    \hfill
    \begin{minipage}{0.3\textwidth}
        \centering
        \includegraphics[width=\linewidth]{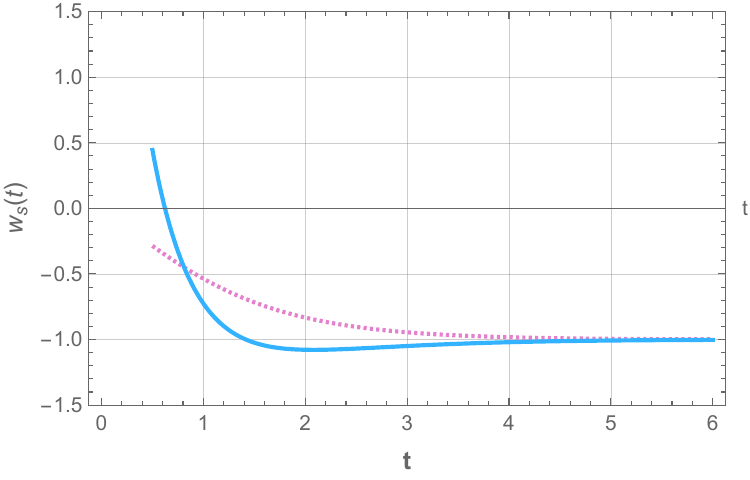}
    \end{minipage}
    \hfill
    \begin{minipage}{0.3\textwidth}
        \centering
        \includegraphics[width=\linewidth]{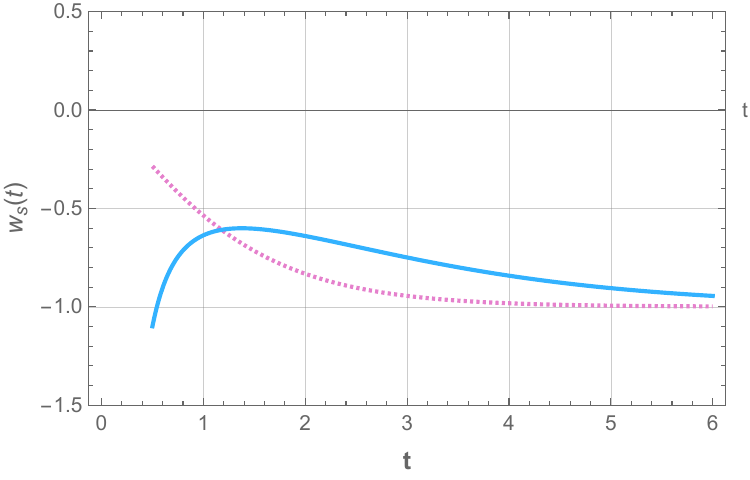}
    \end{minipage}

    \caption{\textbf{(Jordan frame)} Here, the first graph corresponds to $w=1$, the second to $w=1/3$ and last to $w=-1/2$. All parameters are the same, apart from $M_p$ and $\Lambda$ which are set to zero. }

    \label{fig:mpLL0}
\end{figure}

Finally, for completeness, it is also interesting to consider the limit in which the Planck mass and cosmological constant are vanishing, while keeping $\frac{\lambda}{\Mpl^2}$ finite. Vanishing Planck mass usually indicates a strong coupling regime. However, our this is still weakly coupled in this case, as can be seen from the kinetic terms of the scalar and tensor modes. At the same time, this limit is interesting because the action loses the standard Ricci scalar of Einstein's gravity.
By setting the previous conventions and remaining parameters the same, with initial conditions $H(t_i)= 0.5$, $\Theta(t_i)= 9$, we find for various equations of state results depicted in Figure \ref{fig:mpLL0}.

\section{The minimal frame}

The previous case of the matter which is non-minimally coupled to the constrained scalar is very intriguing. It holds potential relevance for both early- and late-time Universe Cosmology, and allows the  matter to contribute to the cosmological evolution, which otherwise was not the case. 
Then the question is which frame should be regarded as the physical frame in which observable quantities are directly described. In our case, it should be the frame in which the matter is minimally coupled to the scalar.
In this section, we will consider the realization of this by introducing the minimal frame -- the frame in which the matter is minimally coupled. For this, we start with the action that we have considered in the previous section (\ref{nminconstrscact}) with the simplest case of non-minimal coupling (\ref{simplestnmincou}):
\begin{equation}
    S=\int d^4x\sqrt{-g}\left(\frac{\Mpl^2}{2}(R-2\Lambda)+\sigma^2\left(-\frac{m^2}{2}+\frac{\xi}{2}R+\lambda\frac{L_m}{\Mpl^2}\right)\right),
\end{equation}
and assume that the Lagrangian density for matter is given by
\begin{equation}
     L_m=p 
\end{equation}
where $p$ is the pressure. In order to absorb the constrained scalar that multiplies the matter, we define
\begin{equation}\label{confmin}
   g_{\mu\nu}=\frac{\Mpl\bar{g}_{\mu\nu}}{\sigma}\,.
\end{equation}
By expressing the action in terms of the new metric, we find
\begin{equation}\label{minframeaction}
    \begin{split}
        S=\int d^4x\sqrt{-\bar{g}}\left[\Mpl\left(\frac{\Mpl^2}{2\sigma}+\frac{\xi\sigma}{2}\right)\bar{R}-\frac{\Mpl}{\sigma}\left(\frac{9\xi}{4}-\frac{3\Mpl^2}{4\sigma^2}\right)\partial_{\mu}\sigma\partial^{\mu}\sigma-\frac{\Mpl^4\Lambda }{\sigma^2}-\frac{\Mpl^2m^2}{2}+\lambda L_m
        \right]\,.
    \end{split}
\end{equation}
We can notice that the term proportional to the Planck mass square in the coefficient of the kinetic term has a wrong sign, implying that the field might become a ghost if this term dominates. 
This is interesting, as the Einstein frame analysis never indicates the presence of a ghost degree of freedom. 
However, we should keep in mind that the theory also has non-minimal coupling, which affects the counting of the degrees of freedom. 
In particular, after performing the analysis, one should find a result that matches the number and type of the degrees of freedom found in the Jordan or Einstein frame. In other words, the theory has a healthy scalar mode.
Another notable feature is that the mass term has effectively become the cosmological constant in the minimal frame.

\subsection{Equations and numerical solutions}
Let us now consider the dynamics in this frame, by studying the equations of motion. As before, we will define $\bar{p}=\lambda p/\Mpl^2$ and $\bar{\varepsilon}=\lambda \varepsilon/\Mpl^2$. 
First, the Hamiltonian constraint is found as
\begin{equation}
    3\left(\frac{\Mpl^{2}+\xi\sigma^2}{\sigma}\right) H^{2}+3\left( \frac{\xi\sigma^2-\Mpl^{2}}{\sigma^{2}}\right) \dot{\sigma} H -\frac{3}{4}\left(\frac{3\xi\sigma^2-\Mpl^2}{\sigma^{3}}\right) \dot{\sigma}^{2}-\Mpl\left(\bar{\varepsilon}-\frac{m^{2}}{2}-\frac{\Mpl^{2} \Lambda}{\sigma^{2}}\right)=0\,.
    \label{minHconstr}
\end{equation}
The variation with respect to the scale factor $a$ gives the acceleration equation, 
\begin{equation}
    \begin{split}
        &-8\sigma^2\left(\xi  \,\sigma^{2}+\Mpl^{2}\right) \ddot{a}-4\sigma^2\left(\xi  \sigma^{2}+\Mpl^{2}\right)a H^{2}-8\sigma\dot\sigma\left(\xi \sigma^{2}- \Mpl^{2} \right)a H\\
        & -4 \left(\xi  \sigma^{2}-\Mpl^{2} \right)a\sigma\ddot\sigma -\left(9 \xi \sigma^{2}+5 \Mpl^{2} \right)a \dot{\sigma}^{2}+2 \left(m^2-2\bar{p}\right)\Mpl\,a \sigma^{3}+4\Mpl^{3} \Lambda \, a \sigma=0\,,
    \end{split}
\end{equation}
while that with respect to the scalar field $\sigma$ gives 
\begin{equation}
    \begin{split}
        &12\sigma^2\left(\xi  \,\sigma^{2}-\Mpl^{2}\right) \ddot{a}+12\sigma^2\left(\xi  \sigma^{2}-\Mpl^{2}\right)a H^{2}+9\left(-6 \xi  \sigma^2+2\Mpl^{2}\right)\sigma \dot{\sigma}a H \\
        &+6\left(-3\xi \sigma^{2}+ \Mpl^{2}\right)a\sigma \ddot{\sigma}+9\left(\xi \sigma^{2}-\Mpl^{2} \right)a \dot{\sigma}^{2}+8\Mpl^{3} \Lambda  a \sigma
=0\,.
    \end{split}
\end{equation}
Finally, the matter energy conservation equation is given simply by 
\begin{equation}
    \dot{\varepsilon}+3H(\varepsilon+p)=0\,.
\end{equation}

Our task is to determine the evolution of the scale factor and the scalar.
From the energy conservation equation, assuming $p=w\varepsilon$ with a constant $w$, we have
\begin{equation}
    \varepsilon=\varepsilon_0\left(\frac{a_0}{a}\right)^{3(1+w)}\,.
    \label{energycons}
\end{equation}
Next, noting that both the acceleration equation and the scalar field equation involve second time derivatives of $a$ and $\sigma$, we may solve for $\ddot{a}$ and $\ddot{\sigma}$ to obtain
\begin{align}
\ddot{a}=&-\frac{a \left(5 \xi^{2} \sigma^{4}-2 \Mpl^{2} \xi  \,\sigma^{2}+\Mpl^{4}\right) H^{2}}{8 \sigma^{4} \xi^{2}}+\frac{a \left(3 \xi^{2} \sigma^{4}-4 \Mpl^{2} \xi  \,\sigma^{2}+\Mpl^{4}\right) \dot{\sigma} H}{8 \sigma^{5} \xi^{2}}
  \nonumber\\&
 -\frac{a \left(33 \xi^{2} \sigma^{4}-6 \Mpl^{2} \xi  \,\sigma^{2}+\Mpl^{4}\right) \dot{\sigma}^{2}}{32 \xi^{2} \sigma^{6}}
 \nonumber\\&
 -\frac{a \Mpl}{16 \sigma^{5} \xi^{2}} \left(-3 \xi  \left(m^{2}-2 \bar{p}\right) \sigma^{4}+\Mpl^{2} \left(m^{2}-\frac{10 \Lambda \xi}{3}-2 \bar{p}\right) \sigma^{2}-\frac{2 \Lambda \,\Mpl^{4}}{3}\right)\,,
\label{minddota}
\\
~\nonumber\\
 \ddot{\sigma}=&-\frac{\left(-\xi^{2} \sigma^{4}+\Mpl^{4}\right) H^{2}}{4 \sigma^{3} \xi^{2}}-\frac{\left(11 \xi^{2} \sigma^{4}+2 \Mpl^{2} \xi  \,\sigma^{2}-\Mpl^{4}\right) \dot{\sigma} H}{4 \xi^{2} \sigma^{4}}-\frac{\left(3 \xi^{2} \sigma^{4}-4 \Mpl^{2} \xi  \,\sigma^{2}+\Mpl^{4}\right) \dot{\sigma}^{2}}{16 \xi^{2} \sigma^{5}}
 \nonumber\\&
 -\Mpl\frac{-3 \xi  \left(m^{2}-2 \bar{p}\right) \sigma^{4}+ \left(3m^{2}-{14 \Lambda \xi}-6 \bar{p}\right) \Mpl^{2} \sigma^{2}-2 \Lambda \,\Mpl^{4}}{24 \sigma^{4} \xi^{2}}\,.
   \label{minddotsig}
\end{align}
Notably, the equation for $\sigma$ simplifies if one substitutes the constraint equation \eqref{minHconstr} for $H^2$,
\begin{equation}
    \ddot{\sigma}+3 \dot{\sigma} H=\frac{\Mpl \left(2 m^{2}+\bar{\varepsilon}-3 \bar{p}\right)}{12 \xi}-\frac{\Mpl^{3} \left(-8 \Lambda \xi +2 m^{2}+\bar{\varepsilon}-3 \bar{p}\right)}{12 \xi^{2}\sigma^{2}}\,.
    \label{minsigeq}
\end{equation}

\begin{figure}[h]
    \centering

    \begin{minipage}{0.45\textwidth}
        \centering
        \includegraphics[width=\linewidth]{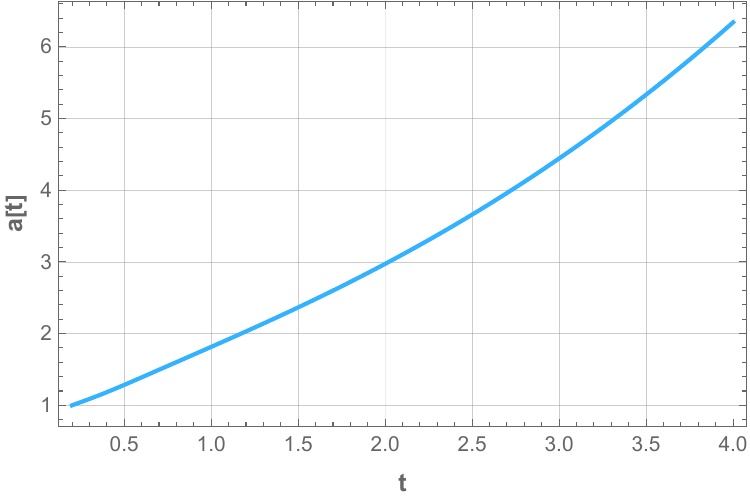}
    \end{minipage}
    \hfill
    \begin{minipage}{0.45\textwidth}
        \centering
        \includegraphics[width=\linewidth]{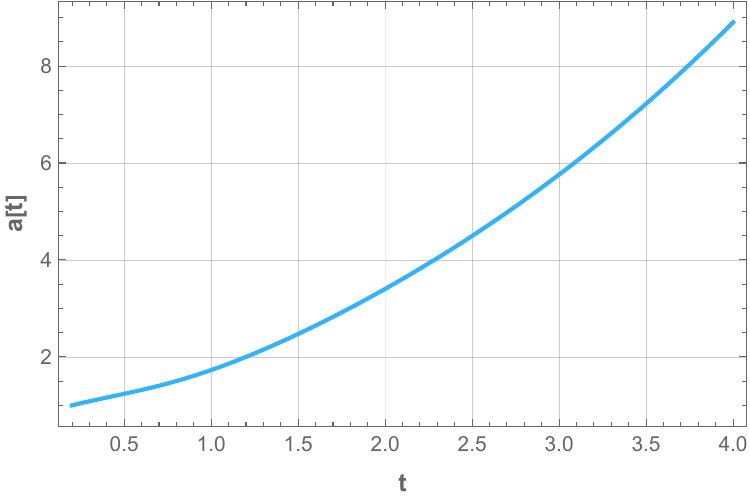}
    \end{minipage}
\vspace{5mm}
    \centering

    \begin{minipage}{0.45\textwidth}
        \centering
        \includegraphics[width=\linewidth]{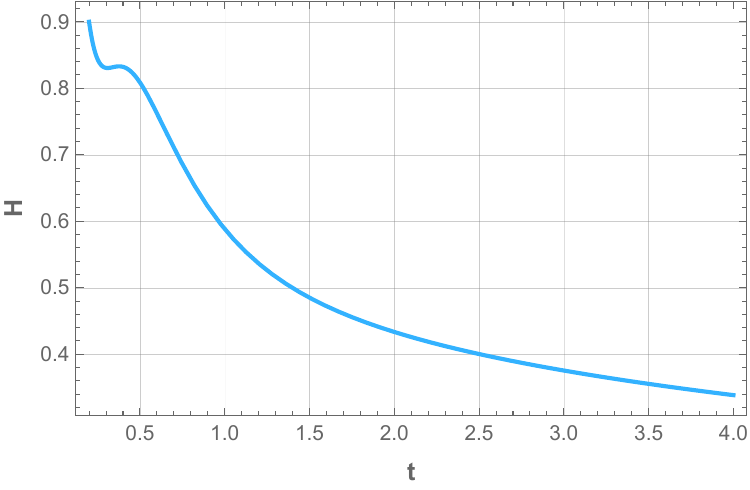}
    \end{minipage}
    \hfill
    \begin{minipage}{0.45\textwidth}
        \centering
        \includegraphics[width=\linewidth]{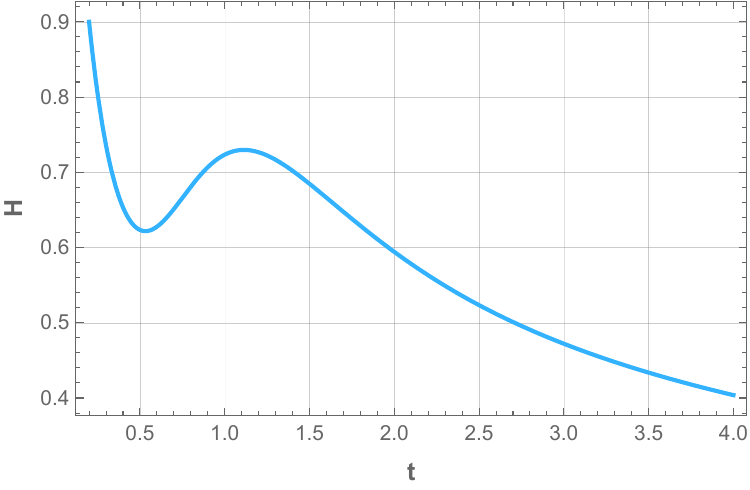}
    \end{minipage}

    \caption{\textbf{(The minimal frame)} The evolution of the scale factor (Upper pannel) and the Hubble parameter (Lower pannel) in the minimal frame. 
    The left ones correspond to $w=1$ and the right ones to $w=1/3$.  }
 \label{fig:minscalefac}

\end{figure}

\begin{figure}[h]
    \centering

    \begin{minipage}{0.45\textwidth}
        \centering
        \includegraphics[width=\linewidth]{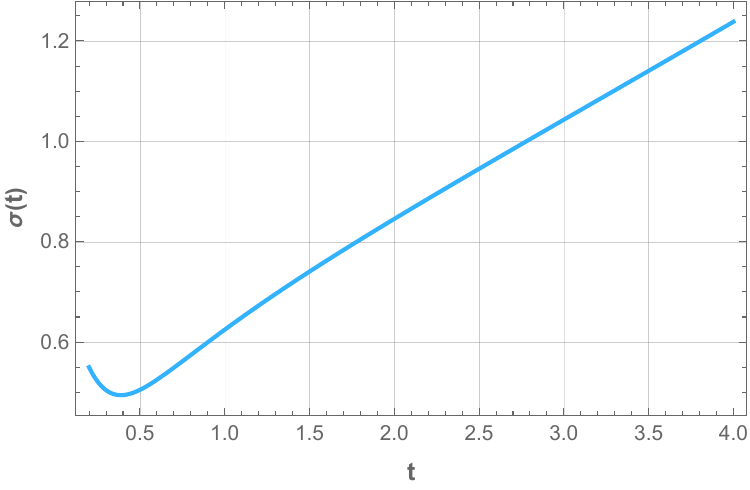}
    \end{minipage}
    \hfill
    \begin{minipage}{0.45\textwidth}
        \centering
        \includegraphics[width=\linewidth]{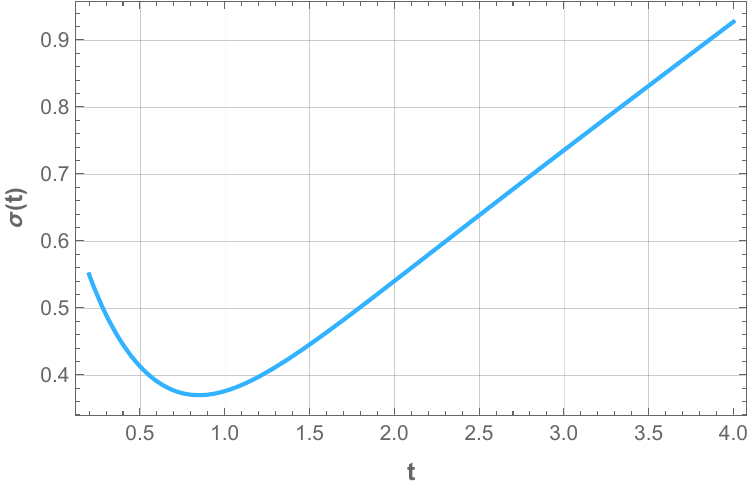}
    \end{minipage}

\vspace{5mm}
    \centering

    \begin{minipage}{0.45\textwidth}
        \centering
        \includegraphics[width=\linewidth]{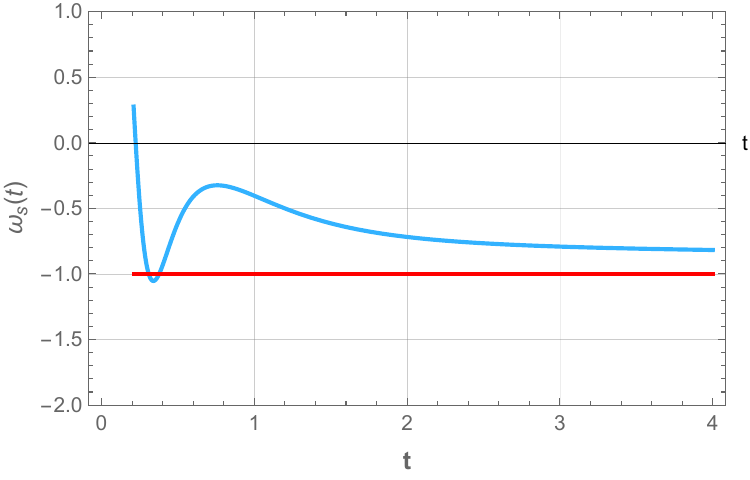}
    \end{minipage}
    \hfill
    \begin{minipage}{0.45\textwidth}
        \centering
        \includegraphics[width=\linewidth]{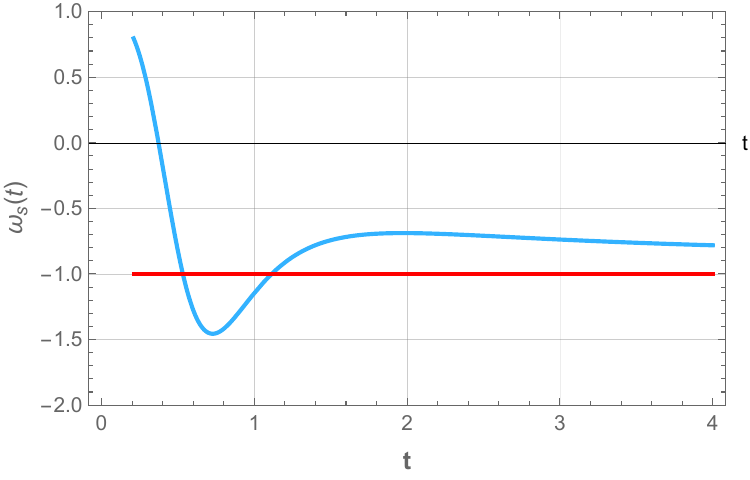}
    \end{minipage}
    \caption{ \textbf{(The minimal frame) } The evolution of the scalar field (Upper pannel) and the effective equation of state (Lower pannel) in the minimal frame. The left ones correspond to $w=1$ and the right ones to $w=1/3$. The red horizontal lines in the lower pannel indicate $w_S=-1$, below which the expansion is in a phantom phase.  }
  \label{fig:minscalar+eos}
\end{figure}

We numerically solve \eqref{minsigeq} and \eqref{minddota} for several cases of the equation of state.
We set the parameters to 
\begin{equation}
    m=1\,,\quad\quad\xi=1\,,\quad \Lambda=1/3\,,\quad \lambda=1
\end{equation}
in the reduced Planck mass units $\Mpl=1$, and choose the initial conditions for $\sigma$, $\dot\sigma$, $a$, and $\dot a$. 
Then, we determine the initial energy density $\varepsilon_0$ to satisfy the Hamiltonian constraint \eqref{minHconstr}. 
At this point, we are interested in the initial conditions that might give rise to the phantom-like effective equation of state.  
By trial and error, we found that it can be achieved by the initial conditions,
\begin{equation}
   H(t_i)=0.9\,,\quad\sigma(t_i)=0.55\,,\quad\dot{\sigma}(t_i)=-0.7\,,
\end{equation}
for both $w=1/3$ and $w=1$. 
This corresponds to $\varepsilon_0=8.72$.
The results are presented in Fig.~\ref{fig:minscalefac} for $a$ and $H$, and in Fig. \ref{fig:minscalar+eos} for $\sigma$ and $w_S$.  

One should take the above results as examples of solutions with a phantom-like effective equation of state. 
In addition, one can obtain similar behavior for dust, or a negative equation of state, for different initial conditions. 
One such example corresponds to setting 
\begin{equation}
    a(t_i)=1\,,\quad H(t_i)=0.3\,,\quad\sigma(t_i)=0.5\,,\quad\dot{\sigma}=-0.4\,,
\end{equation}
among other choices. 
The solutions without a dip in the Hubble parameter are possible as well, in which case the effective equation of state $w_s$ is always larger than $-1$.
A set of initial conditions that recovers this type of solutions is given by
\begin{equation}
    H(t_i)=1.2\,,\quad\sigma(t_i)=5\,,\quad\dot{\sigma}(t_i)=0.4\,.
\end{equation}

\subsection{Connection to the Jordan frame}
\begin{figure}[h!]
    \centering

    \begin{minipage}{0.3\textwidth}
        \centering
        \includegraphics[width=\linewidth]{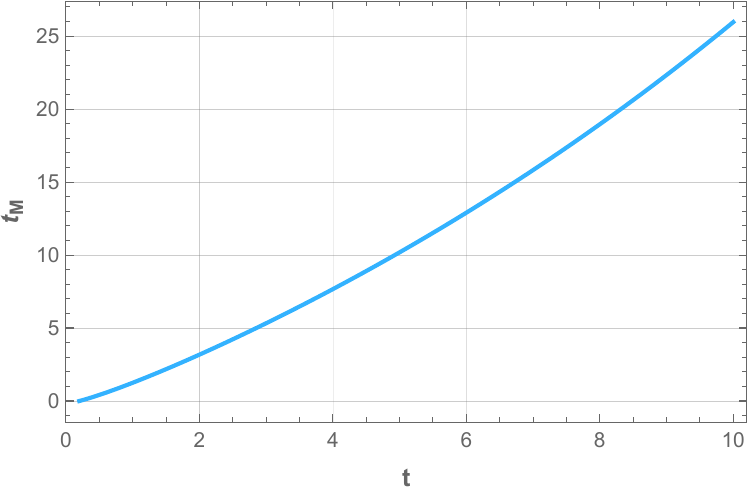 }
    \end{minipage}
    \hfill
    \begin{minipage}{0.3\textwidth}
        \centering
        \includegraphics[width=\linewidth]{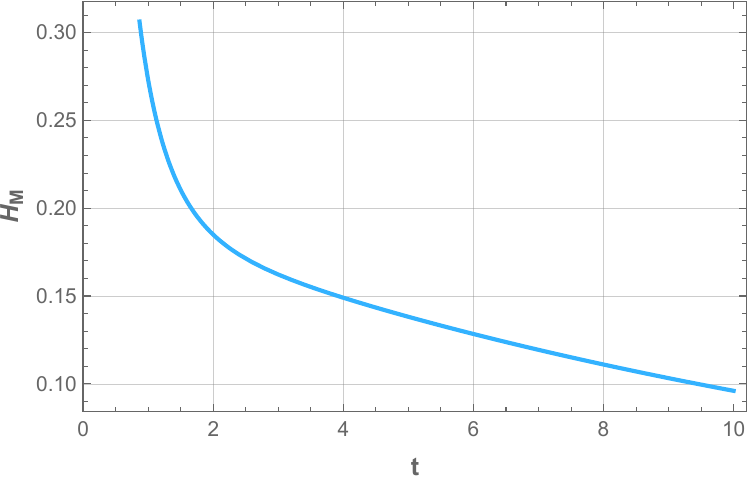 }
    \end{minipage}
    \hfill
    \begin{minipage}{0.3\textwidth}
        \centering
        \includegraphics[width=\linewidth]{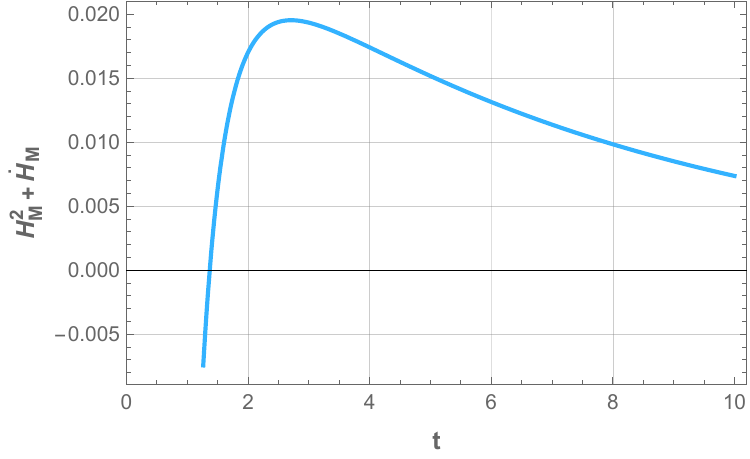 }
    \end{minipage}
    \center
    \begin{minipage}{0.3\textwidth}
        \centering
        \includegraphics[width=\linewidth]{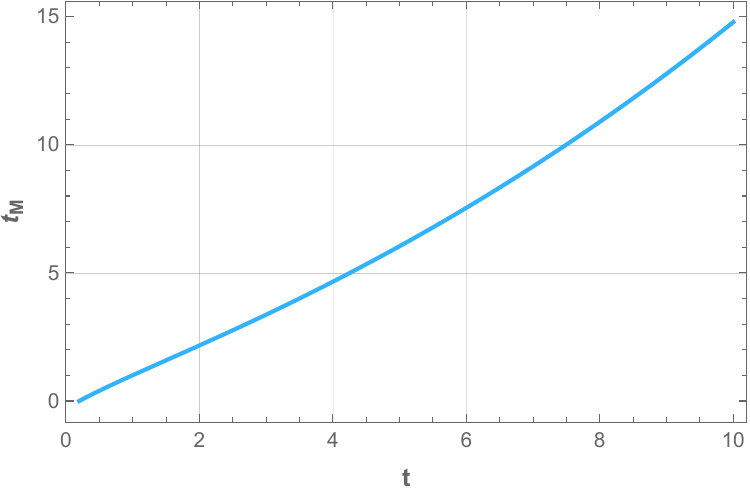 }
    \end{minipage}
     \hfill
    \begin{minipage}{0.3\textwidth}
        \centering
        \includegraphics[width=\linewidth]{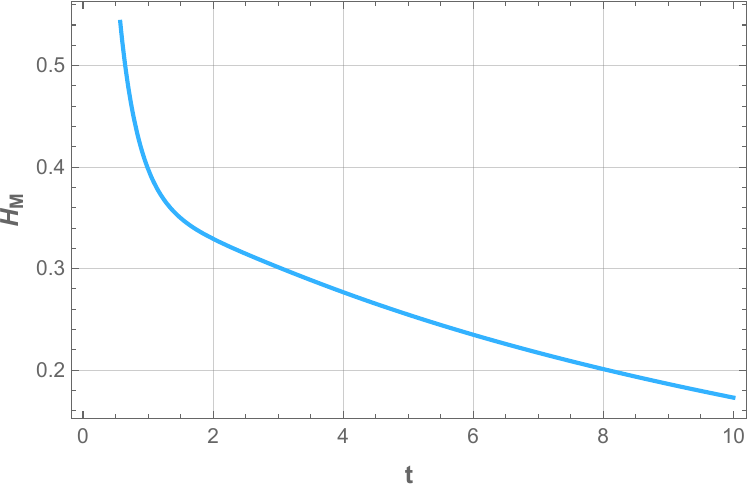 }
    \end{minipage}
     \hfill
    \begin{minipage}{0.3\textwidth}
        \centering
        \includegraphics[width=\linewidth]{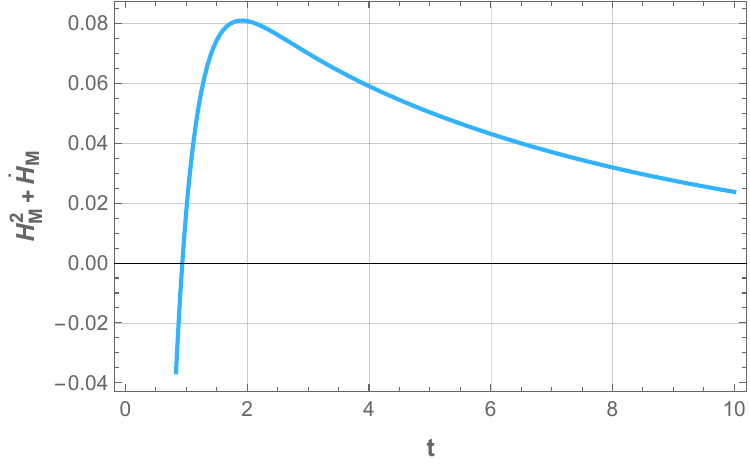 }
    \end{minipage}
    \center
    \caption{ \textbf{(The minimal frame) }{ The relation between the Minimal frame, and the solution that describes the equation of state for radiation $w=1/3$ in the Jordan frame with two cases of initial conditions.}
\\
    {\bf Upper panel:}{ This case corresponds to the Jordan-frame solution given in the last plot in Figure \ref{fig:superH}. On the left, the minimal time is expressed in terms of the Jordan time. In the centre, we have shown the decrease in the Hubble parameter, while the last plot shows acceleration.  }  
    \\
  {\bf Lower panel:}{ This case corresponds to the solution depicted in the Figure \ref{fig:radiation} in the Jordan frame. The first plot depicts the time in the minimal frame as a function on that given in the Jordan frame, while the seconds plot depicts the Hubble parameter. The last plot shows acceleration.  } 
  We can notice that both upper and left panel contain a transition from decelerated to accelerated expansion. 
}\label{finalpic}
\end{figure}

Let us now consider the connection between the minimal and the Jordan frame. Similarly to the Einstein frame analysis, the metric in the minimal frame, $\bar{g}_{\mu\nu}$ is connected to the Jordan frame one by a conformal transformation. Let us denote the time that corresponds to the minimal frame with $t_M$, and reserve the time $t$ for the Jordan-frame one. The line element in the minimal frame is
\begin{equation}
  \begin{split}
        ds_M^2&=\sigma ds_J^2=\sigma g_{\mu\nu}dx^{\mu}dx^{\nu}=\sigma(-dt^2+a^2\delta_{ij}dx^idx^j)\\
        &=-dt_M^2+a^2_M\delta_{ij}dx^i_Mdx^j_M. 
  \end{split},
\end{equation}
From the above two relations, we establish 
\begin{equation}
    dt_M=\sqrt{\sigma}dt\qquad a_M=\sqrt{\sigma}a\qquad x_M=x\,.
\end{equation}
Similarly to the Einstein frame, the Hubble parameter in the minimal frame is related to the one in the Jordan frame by 
\begin{equation}
    H_M=\frac{1}{a_M}\frac{d a_M}{dt_M}= \frac{\dot{\sigma}}{2\sigma^{3/2}}+\frac{H}{\sqrt{\sigma}}\,.
\end{equation}
Let us consider two different examples, focusing on the equation of state that corresponds to radiation, $w=1/3$. In the first, case, we will set the initial conditions to match the case with super-Hubble expansion in the Jordan frame, given in the last plot in Fig.~\ref{fig:superH}. In the second case, we will consider the case given in the Fig.~\ref{fig:radiation}. Then, by using the above relations, we find the results that are depicted in the Fig.~\ref{finalpic}. Both solutions show a transition from a decelerated to an accelerated phase.

\section{Discussion}
The substances that underline the early or late-time acceleration of the Universe are some of the greatest mysteries that we have yet to uncover. While many models exist, the latest observations urge us to explore theories beyond the known ones and discover different mechanisms that could give rise to the dynamics of the Universe that we see today.
 
In this work, we have studied one of such mechanisms, which is based on the notion of constraints -- an essential part of any gauge theory, that involves fields without a standard kinetic term. 
In our case, we have considered a framework with a scalar field without a kinetic term that gives rise to a cosmological background that is independent of the cosmological constant.
Curiously, in the simplest version that involves only the coupling between the scalar field and the Ricci scalar, we have found potential for our framework to give rise to cosmologically interesting solutions for the early- and late-time acceleration of our Universe. 
Although the models studied in this paper has nothing to do with the Universe we observe today, they give a hint to uncovering both the cosmological constant problem and substance that drives the acceleration of the Universe. 
 
Throughout this work, by focusing on the simplest realization of the constrained scalar field framework, we have considered a theory consisting of two parts -- the first including the standard Einstein gravity together with a cosmological constant, and the second involving a constrained scalar, a field with only a mass term, which is coupled non-minimally to the Ricci scalar. Due to the presence of constraints, we have seen that this theory has two key branches. In the first branch, the constrained scalar can be set to zero, in which case the underlying background obeys standard cosmology.
Notably, in the second branch which involves a non-vanishing field, the constraint entirely determines the values of the Ricci scalar. This results in the evolution of space-time which is independent of the values of the cosmological constant, but is determined by the values of the mass and coupling constant of the scalar field. 
The cosmological constant, on the other hand, contributes only to the evolution of the background scalar.  
 
By studying the solutions in this second branch of the theory, we have found several analytical solutions that yield interesting dynamics of space-time. In particular, in the Jordan frame, we have found that the Universe can transition from a radiation-dominated phase to an exponential expanding phase, even in the case of a free theory without matter. 
In addition, the theory can also be relevant for the early-universe cosmology. Starting initially from a strongly coupled regime, the Universe can undergo a super-Hubble expansion -- a stage in which the Hubble parameter increases with time -- which later on relaxes to an exponential stage with a constant Hubble parameter. One should note that similar super-Hubble expansion can also be found in Proca theory with non-minimal coupling to gravity \cite{DeFelice:2025ykh}. However, this theory faces several challenges including the appearance of instabilities in the anisotropic case. In contrast, by studying the perturbations of our model in the Jordan frame, we have shown that the theory contains a well-behaved scalar mode and two tensor modes. This is further supported in the Einstein frame, where the theory can be written in terms of Einstein gravity, and a scalar field with a potential. 
 
Due to the peculiar structure of the theory, on the one hand, we have seen that matter with minimal coupling to both scalar field and gravity, can only contribute to the background dynamics of space-time if the constrained scalar is vanishing. On the other hand, if the external matter is non-minimally coupled to the constrained scalar, it can give rise to interesting behavior, which includes also the phantom crossing of the equation of state. One can nevertheless wonder if this is only an artifact of the Jordan frame, or could give rise to such behavior in the physical frame as well. To explore it further, we have introduced a minimal frame -- frame in which the matter is minimally coupled to the constrained scalar, which plays the role of the physical frame, and where we have showed that phantom-like crossing may take place as well.
One should note that in the case of minimal coupling between the matter and the constrained scalar with $\sigma=0$, one recovers the standard coupling between the Einstein gravity and the energy-momentum tensor of the matter. In future studies, it would be interesting to explore if the $\Lambda$CDM case (including the case of a time-dependent dark energy) could also be reproduced with the non-minimal coupling between matter and the scalar, and investigate also other cases, such as the coupling between the trace of the energy-momentum tensor and the constrained scalar.

To conclude, by considering a simple model of a scalar field with non-minimal coupling to gravity and without kinetic term, we found two interesting features: The independence of the cosmological evolution on the values of the cosmological constant, and the existence of a mechanism that gives rise to the accelerating universe without cosmological constant. 
While the model we considered is still far from a realistic cosmological scenario, {the properties that we have found by studying the simplest case of this theoretical framework of the constrained scalar indicate that a cosmologically viable solution that matches with observation may be found upon further extensions, and investigations of the predictions of this theory.  Moreover, it would also be  } of interest to test it against the current observational constraints, involving, especially the PPN parameters \cite{Chiba:1997ij}.
{The theory of a constrained scalar that we have analysed in this work} serves as a basis for the construction of many interesting extensions, ranging from scalar and vector fields or even higher curvature terms, all based on the notion of constraint -- a non-propagating field, which can be found in any gauge theory. 

\begin{center}
\textbf{Acknowledgments}
\end{center}

%\textit{
A. H. would like to thank Antonio De Felice for the initial involvement in this topic, Dieter L\"ust, Michael Haack, Viatcheslav Mukhanov and Jun'ichi Yokoyama for useful discussions, and Elisa G. M. Ferreira for useful comments. 
A. H. also thanks LMU Munich and RBI Zagreb for hospitality during her visit, when part of this work was carried out. 
This work was supported in part by JSPS KAKENHI No.~JP24K00624.
The work of A. H. was supported in part by the JSPS KAKENHI Grant No. JP26K17133, the
CD3 Google Seed grant, and
by the World Premier International Research Center Initiative (WPI), MEXT, Japan. 
%}

\bibliographystyle{utphys}
\bibliography{paper}{}

\end{document}